\documentclass[prd,preprintnumbers,nofootinbib]{revtex4} 
\usepackage{graphicx} 
\usepackage{amsmath}
\usepackage{amsfonts,amsbsy}
\usepackage{amssymb}
\usepackage{appendix}

\def\k{{\mathbf k}}
\def\q{{\mathbf q}}
\def\b{{\mathbf b}}
\def\r{{\mathbf r}}

\def\be{\begin{equation}}
\def\ee{\end{equation}}
\def\bea{\begin{eqnarray}}
\def\eea{\end{eqnarray}}

\def\Tr{{\rm tr}}
\def\eq#1{{Eq.~(\ref{#1})}}
\def\fig#1{{Fig.~\ref{#1}}}

\begin{document}

\title{\bf Photon-jet ridge at RHIC and the LHC  }


\author{Amir H. Rezaeian}
\affiliation{
 Departamento de F\'\i sica, Universidad T\'ecnica
Federico Santa Mar\'\i a, Avda. Espa\~na 1680,
Casilla 110-V, Valparaiso, Chile\\
Centro Cient\'\i fico Tecnol\'ogico de Valpara\'\i so (CCTVal), Universidad T\'ecnica
Federico Santa Mar\'\i a, Valpara\'\i so, Chile}
\date{\today}

\begin{abstract}
We investigate long range rapidity correlations of pairs of prompt photon and jet in the Color Glass Condensate (CGC) framework in  proton-proton and proton-nucleus collisions at RHIC and the LHC. We show that photon-jet correlations exhibit long-range azimuthal collimation at near-side for low transverse momenta of the produced photon and jet in high-multiplicity events. These ridge-like features are strikingly similar to the observed ridge effect for di-hadron correlations at RHIC and the LHC. We show that  correlations in the relative rapidity and the relative azimuthal angle between pairs of prompt photon and jet strongly depend on the gluon saturation dynamics at small-x kinematics and such  measurements can help to understand the true origin of the observed di-hadron ridge in p+A collisions, and address whether the ridge is a universal  phenomenon for all two particle correlations at high energy and high multiplicity events. We also investigate if there is a ridge-like structure for photon-hadron pair correlations at RHIC and the LHC. We found that the hadronization of jet has non-trivial effects on the photon-jet correlations.  
\end{abstract}

\maketitle

\section{Introduction}
Two particle correlations in high-energy collisions have played significant role to unravel QCD novel phenomena \cite{exp-pp,exp-pa1,exp-pa2,exp-pa3,exp-pa4,exp-pa5,exp-pa6,novel,di-e,ph-ex}. 
The recent discovery of the so-called ridge phenomenon, namely di-hadron collimation in the relative azimuthal angle $\Delta \phi=\phi_1-\phi_2$ between pairs at near-side $\Delta \phi\approx 0$ (and at away-side $\Delta \phi\approx \pi$) which extend to a large pseudorapidity separation between pairs  $\Delta\eta$, in high-multiplicity events selection in both proton-proton and proton(deuteron)-nucleus collisions at the LHC and RHIC 
\cite{exp-pp,exp-pa1,exp-pa2,exp-pa3,exp-pa4,exp-pa5,exp-pa6} and more recently in $^3\text{He+Au}$ collisions at RHIC \cite{he-au}, triggered an on-going debate about the underlying dynamics of high-multiplicity events in small colliding systems.  
Strikingly, such correlations when triggering on rare events with high multiplicities (not present in minimum bias proton-proton collisions) show great deal of similarity to that measured in semi peripheral nucleus-nucleus collisions \cite{exp-hic}. The observation of a ridge in high-multiplicity p+Pb collisions 
\cite{exp-pa1,exp-pa2,exp-pa3,exp-pa4,exp-pa5,exp-pa6}
was less surprising compared to the same effect in proton-proton (p+p) collisions \cite{exp-pp}, but it was unexpected that the strength of the correlation is almost as large as the ridge in heavy-ion collisions. The ridge-type structure in nucleus-nucleus (A+A) collisions at RHIC and the LHC is fairly understood as a phenomenon related to hydrodynamical behavior of the quark-gluon-plasma, see for example Refs.\,\cite{exp-geo1,exp-geo2}. However, it is still an outstanding problem to understand how (and why) a small system like the one produced in p+p collisions, an order of magnitude  smaller than in heavy-ion collisions, exhibits a hydro type behavior \cite{hydro-1,hydro-2}.

The initial-state Color-Glass-Condensate (CGC) effective field theory approach provides an alternative description of the ridge effect in small colliding systems \cite{ridge1,ridge2,ridge3,ridge4,ridge5}. Within the initial-state CGC framework, it was shown that the ridge effect can arise from three rather different mechanisms: Glasma graphs \cite{ridge1,ridge4}, local anisotropy of target fields  \cite{ridge2} and spatial variation of partonic density  \cite{ridge3}. In principle, all these three mechanisms can be relevant to understanding the true physical origin of the ridge in p+p(A) collisions \cite{ridge-rev1}. Note that it was recently shown that the calculation based on the Glasma graphs \cite{ridge1,ridge4} can be understood in terms of Bose enhancement of gluons in the hadronic wave-functions \cite{ridge5}. A complete analysis of the observed ridge effect in p+p(A) collisions which incorporates all above-mentioned contributions is still missing. For recent reviews on the ridge effect, see Refs.\,\cite{ridge-rev1,ridge-rev2,ridge-rev3}. 

 It is currently unknown whether the ridge is a universal  phenomenon, for all different two-particle productions such as di-hadron, photon-jet, di-jet and di-photon production in p+p(A) collisions, see Ref.\,\cite{di-photon}. In this paper, for the first time, we investigate whether based on the initial-state physics there is a ridge-type structure for prompt photon and jet pair production in high multiplicity p+p(A) collisions at RHIC and the LHC.  
We show near-side collimation of photon-jet production in p+p(A) collisions strongly depends on the gluon saturation dynamics and such measurements can help to understand the true origin of ridge effect in high-multiplicity events.

Photons radiated in hard collisions not via hadronic decays are usually called prompt photon. There are advantages to studying prompt photon and jet pair production as compared to di-hadron (and di-jet) production. It is theoretically cleaner; one avoids the difficulties involved with description of hadronization and possible initial-state-final-state interference effects which may be present for hadrons pair production. 
In particular, in the inclusive di-jet production at leading-order approximation, higher number of Wilson lines, the Weizs\"acker-Williams and the dipole gluon distributions are involved \cite{di-1,di-2,di-3,di-4,di-5} while in diffractive DIS di-jet \cite{dijet-diff1,dijet-diff2}, photon-jet \cite{pho-cgc1,pho-cgc2,ja,ja2} and prompt di-photon \cite{di-photon} cross-sections, only dipole gluon distribution appears which is both experimentally and theoretically well-known, see for example Refs.\,\cite{jav1,ip-sat,bcgc}. Moreover,  in contrast to gluons, prompt photons do not scatter on the target gluon field, but rather decohere from the projectile wavefunction due to scatterings of the quarks. Therefore,
the underlying mechanism of prompt photon and quark pair production is quite different from two gluons (and quark-gluon) pair production, and such measurements  can provide vital complementary information to address whether it is the initial or final state effects that play dominant role in formation of the ridge collimation in p+p(A) collisions.  Note that the photon-hadron (and photon-jet) correlations have also been a very powerful probe of the in-medium parton energy loss in high-energy heavy-ion collisions \cite{ph-ex,ph-th}.

It  was recently shown \cite{ja,ja2} that the correlation of the back-to-back photon-hadron pair production in high-energy p+p and p+A collisions can be used to probe the gluon saturation at small-x region and to study the physics of cold nuclear matter in dense region. 
However, in the previous studies, photon-hadron correlations were only studied in a small kinematic window (concentrating mostly in away-side  correlations at $\Delta\phi\approx \pi$) in minimum bias (or medium impact parameters) collisions. Here, we extend the previous CGC studies, and analysis in details photon-jet and photon-hadron pair correlations in p+p and p+A collisions at finite rapidity separations between prompt photon and jet $\Delta\eta^{\gamma-jet}\ne 0$ for arbitrary azimuthal angle  $\Delta\phi$ between the pairs in high multiplicity events. We show that due to gluon saturation dynamics the photon-jet pairs within forward rapidity separation about  $2\le \Delta\eta^{\gamma-jet}\le 4$ are collimated in the relative azimuthal angle between pairs at near-side $\Delta\phi\approx 0$  at low transverse momenta of pairs in high multiplicity collisions.
These ridge-like features are strikingly similar to the observed ridge effect for di-hadron correlations \cite{exp-pp,exp-pa1,exp-pa2,exp-pa3,exp-pa4,exp-pa5,exp-pa6} at RHIC and the LHC. We provide various predictions for photon-jet correlations (and the ridge)  which can be tested at RHIC and the LHC. The asymmetric nature of photon-hadron (and photon-jet) production, and the fact that in semi-inclusive photon-hadron production, QCD and electromagnetic interactions are inextricably intertwined, make correlations very intriguing. We show that the hadronization of jet has non-trivial effects on the photon-jet correlations.

This paper is organized as follows; In Sec. II, we first provide a concise description of theoretical framework by introducing the main formulas for the calculation of the cross sections of photon-jet and photon-hadron pair production within the CGC approach. In Sec. II we also introduce the observables that we are interested to compute and our numerical setup. In Sec. III, we present our detailed results and predictions. We summarize our main results in Sec. IV.


\section{Main formulations and numerical setup}

It is generally believed that a system of gluons at high energy or small Bjorken-x forms a new state of matter where the gluon distribution saturates \cite{sg}. Such a system is endowed with a new dynamical momentum scale, the so-called saturation scale 
at which non-linear gluons recombination effects become as important as the gluon radiation. The saturation scale controls the general features of bulk of particles produced in small-x region. The color glass condensate (CGC) approach has been proposed to study the physics of gluon saturation at small-x region \cite{mv,bk,jimwlk}. The CGC formalism is an effective perturbative QCD theory in which one systematically re-sums quantum corrections which are enhanced by large logarithms of 1/x and also incorporates density effects.  This framework has been successfully applied to many QCD processes from HERA \cite{jav1,ip-sat,bcgc,na-sat} to RHIC and the LHC \cite{e-lhc,m1,tr,tr1,j1,me-pa}.  For reviews see Refs.\,\cite{cgc-review1,cgc-review2}.

The cross-section for production of a quark and prompt photon in dilute-dense scatterings such as proton-nucleus collisions was obtained within a hybrid approximation in the CGC framework by three independent groups of Gelis-Jalilian-Marian \cite{pho-cgc1}, Baier-Mueller-Schif \cite{pho-cgc2} and Kovner-Rezaeian \cite{di-photon}. In the hybrid approximation \cite{hybrid}, one treats the projectile proton in the parton model and the target nucleus in the Color-Glass-Condensate approach.  Note also that Bremsstrahlung of a quark propagating through a nucleus at the classical limit was first calculated by Kopeliovich-Tarasov-Sch\"afer \cite{boris} in coordinate space using the dipole formalism, see also Refs.\,\cite{me2-pho}. Although the formulations employed in Refs.\,\cite{pho-cgc1,pho-cgc2,di-photon,boris} are different from each others, they give consistent results. Recently, these calculations were extended to consider diphoton-quark and prompt di-photon production in p+A collisions \cite{di-photon}.  
 
The cross section for production of a quark and a prompt photon with $4$-momenta $q$ and $k$ respectively (both on-shell) in scattering of a on-shell quark with $4$-momentum $p$ on a target (either proton or nucleus) in the
CGC formalism is given by \cite{pho-cgc1,pho-cgc2,di-photon}, 
\bea d\, \sigma &=& 4 e^2\,
  e_q^2\frac{d^3 k}{(2\pi)^3\,2 k^-} \frac{d^3 q}{(2\pi)^3\, 2
  q^-} \frac{1}{2 p^-}(2\pi)\, \delta (p^- - q^- -k^-) \, [(p^-)^2 + (q^-)^2] \bigg[\frac{p\cdot q}{p\cdot k\, q\cdot k} + 
\frac{1}{q\cdot k} - \frac{1}{p\cdot k} \bigg] \nonumber\\
&&
d^2 \b_T\, d^2 \r_T\, e^{i (\q_T + \k_T)\cdot
       \r_T} \, N_F (b_t, r_t, x_g),
\label{cs_gen}
\eea
using the explicit forms of the momenta in the above expression, we obtain 
\bea
&&{d\sigma^{q(p)+A(p_A) \rightarrow q(q)+\gamma(k)+ X}
\over d^2\b_T\, dk_T^2\,d\eta^{\gamma}\, dq_T^2\, d\eta^{jet}\, d\phi} =
{e_q^2\, \alpha_{em} \over \sqrt{2}(2\pi)^3} \, 
{k^-\over  k_T^2 \sqrt{S}} \,
{1 + ({q^-\over p^-})^2 \over
|k^- \, \q_T - q^- \k_T|^2}\nonumber \\
&&\delta [x_q - {q_T \over \sqrt{S}} e^{\eta^{jet}} - {k_T \over \sqrt{S}} e^{\eta^{\gamma}} ] \,
\bigg[ 2 q^- k^-\, \q_T \cdot \k_T + k^- (p^- -k^-)\, q_T^2 + q^- (p^- -q^-)\, k_T^2 \bigg] 
\nonumber \\
&&\int d^2 \r_T \, e^{i (\q_T + \k_T)\cdot\r_T}   \, N_F (b_T, r_T, x_g) ,
\label{cs}
\eea
where $\sqrt{S}$ is the nucleon-nucleon center of mass energy and $x_q$ is the ratio
of the incoming quark to nucleon energies such that $p^-=x_q \, \sqrt{S/2}$. Throughout this paper, two-dimensional vectors in transverse space are denoted by boldface. 
The 
outgoing photon and quark  rapidities  $\eta^{\gamma}$ and $\eta^{jet}$ are defined via $k^-={k_T \over \sqrt{2}} e^{\eta^{\gamma}}$
and $q^-={q_T \over \sqrt{2}} e^{\eta^{jet}}$, respectively.  The relative azimuthal angle between the final state
quark-jet and photon $\Delta \phi$ is defined via $\cos(\Delta \phi) \equiv {\q_T \cdot \k_T \over  q_T k_T}$. 
In Eqs.\,(\ref{cs_gen},\ref{cs}), all the multiple scatterings and small $x$ evolution effects are encoded in the imaginary part of quark-antiquark dipole-target forward scattering amplitude $N_F (b_T, r_T, x_g)$, defined as 
\be
N_F(b_T,r_T,x_g) = {1\over N_c} \, < \Tr [1 - V^{\dagger} (x_T) V (y_T) ] >,  
\label{cs_def}
\ee
where $N_c$ is the number of color. The vector $\b_T\equiv (\vec{x_T} + \vec{y_T})/2$ is the impact parameter of the dipole from the target and $\r_T\equiv \vec{x_T} - \vec{y_T}$ denotes the $q\bar{q}$-dipole transverse separation vector. 
The matrix $V (y_T)$ is a unitary matrix in fundamental representation of $SU(N_c)$ containing the interactions of a quark and the colored glass condensate target. The dipole scattering probability depends on Bjorken $x_g$ via the B-JIMWLK renormalization group equations \cite{jimwlk}. 

In order to relate the above partonic production cross-section to 
proton-nucleus collisions, one needs to convolute the 
partonic cross-section in \eq{cs} with the quark and
antiquark distribution functions of a proton or a deuteron,  
\begin{eqnarray}\label{qg-f}
\frac{d\sigma^{p+ A \rightarrow \gamma (k) + \text{jet}(q)+ X}} {d^2\b_T \, d^2\k_T\,d\eta^{\gamma}\, d^2\q_T \, d\eta^{jet}}&=& 
 \int\, dx_q\, \sum_{\text{quark, antiquark}}
f (x_q,Q^2) \frac{d\sigma^{q+ A \rightarrow \gamma (k) + \text{jet}(q)+ X}}{d^2\b_T \, d^2\k_T\,d\eta^{\gamma}\, d^2\q_T d\eta^{jet}},  
\end{eqnarray}
where $f(x_q,Q^2)$ is the parton distribution function (PDF) of the incoming proton (deuteron) which depends on 
the light-cone momentum fraction $x_q$ and the hard scale $Q$. The light-cone fraction variables $x_g$ and $x_q$ in Eqs.\,(\ref{cs_def},\ref{qg-f}) are related to the prompt photon and final-state jet rapidities, and transverse momenta via
\bea
x_g &=& {1\over \sqrt{S}}\left(k_T \exp(-\eta^{\gamma}) + q_T \exp(-\eta^{jet})\right),  \nonumber\\
x_q&=& \frac{1}{\sqrt{S}}\left(k_T\, \exp(\eta^{\gamma})+ q_T\, \exp(\eta^{jet})\right). \
\label{x-1}
\eea

In order to obtain the cross-section for photon-hadron production from the photon-jet cross-section, in addition we should convolute \eq{qg-f} with the quark-hadron 
fragmentation function,
\begin{eqnarray}\label{hg-f}
\frac{d\sigma^{p+ A \rightarrow \gamma (k) + \text{h} (q^h)+ X}}{d^2\b_T \, d^2\k_T\,d\eta^{\gamma}\, d^2\q^{h}_T \, d\eta^{h}}&=& \int^1_{z_{h}^{min}}\frac{dz_h}{z_h^2} \int\, dx_q\,\sum_{\text{quark, antiquark}} f (x_q,Q^2) \frac{d\sigma^{q+ A \rightarrow \gamma (k) + \text{jet}(q)+ X}}{d^2\b_T \, d^2\k_T\,d\eta^{\gamma}\, d^2\q_T d\eta^{jet}} D_{h/q}(z_h,Q^2),
\end{eqnarray}
where $q^h_T$ is the transverse momentum of the produced hadron and the function $D_{h/q}(z_h,Q)$ is the quark-hadron fragmentation function (FF)  where $z_h$ is the ratio of energies of the produced hadron and quark. We recall that for massless hadrons, rapidity and pseudo-rapidity $\eta$ is the same. Here, we only consider light hadrons, and we make no distinction between the
rapidity of a quark and the hadron to which it fragments. Moreover due to the assumption
of collinear fragmentation of a quark into a hadron, the angle $\Delta \phi$ is now the angle between the produced photon and hadron.
Similar to the case of the photon-jet production, the light-cone momentum fraction variables $x_q$ and $x_g$ in the case of photon-hadron production are related to the transverse momenta and rapidities of the produced hadron (or the produced jet) and prompt photon via energy-momentum conservation \cite{ja}, 
\begin{eqnarray}
x_q&=&\frac{1}{\sqrt{S}}\left(k_T\, \exp(\eta^{\gamma})+\frac{q^h_T}{z_h}\, \exp(\eta^{h})\right),\nonumber\\
x_g&=&\frac{1}{\sqrt{S}}\left(k_T\, \exp(-\eta^{\gamma})+ \frac{q^h_T}{z_h}\, \exp(-\eta^{h})\right),\nonumber\\
z_h&=&q^h_T/q_T \hspace{1 cm} \text{with}~~~~~ 
z_{h}^{min}=\frac{q^h_T}{\sqrt{S}} \left(\frac{\exp(\eta^h)} {1 - {k_T\over \sqrt{S}}\, \exp(\eta^{\gamma})}\, \right). \label{x-2}\
\end{eqnarray}

The single inclusive prompt photon cross section can be readily obtained from
\eq{cs_gen} or \eq{cs} by integrating over the momenta of the final state quark. After shifting $\q_T\to \q_T-\k_T$ we obtain, 
\begin{eqnarray}\label{pho}
\frac{d\sigma^{q+A \rightarrow \gamma (k) + X}}{d^2\b_T dk^2_T d\eta^{\gamma}}&=& 
\frac{e_q^2 \alpha_{em}}{(2\pi)^3} z_\gamma^2[1+(1-z_\gamma)^2]
\frac{1}{k^2_T} \int d^2 \r_T \,d^2 \q_T 
\frac{q_T^2}{[z_\gamma\, \q_T - \k_T ]^2}
\, e^{i \q_T\cdot \r_T}   \, N_F (b_T, r_T, \mathcal{X}_g), 
\end{eqnarray}
where $z_\gamma \equiv k^-/p^-$ denotes the fraction of the projectile quark energy $p^-$ carried by
the photon and $d\eta_\gamma \equiv \frac{d z_\gamma}{z_\gamma}$. Various limits of this expression have 
been studied in \cite{pho-cgc2} where it was shown that in the limit where photon has a large transverse momentum $k_T \gg z_\gamma\, q_T$ such that the collinear singularity is suppressed, one recovers the 
LO pQCD result for direct photon production process $q\, g \rightarrow q\, \gamma$ 
convoluted with the unintegrated gluon distribution function of the target. On the other hand,
if one performs the $q_T$ integration without any restriction, one recovers  
the LO pQCD expression for quark-photon fragmentation function convoluted with 
dipole scattering probability. In the limit where one can ignore multiple scattering of
the quark on the target ("leading twist" kinematics), this expression reduces to the 
pQCD one describing LO production of fragmentation photons.

In order to relate the partonic cross-section given by \eq{pho} to photon production in 
p+A collisions, we convolute \eq{pho} with PDFs of the projectile proton, 
 \begin{equation}\label{pho-f}
\frac{d\sigma^{p+ A \rightarrow \gamma (k) \, X}}{d^2\b_T d^2\k_T d\eta^{\gamma}}=  
\int_{\mathcal{X}_q^{min}}^1 d \mathcal{X}_q \sum_{\text{quark, antiquark}} f_q(\mathcal{X}_q, k_T^2)
\frac{d\sigma^{q(p) + A \rightarrow \gamma (k) + X}}{d^2 \b_T d^2\k_T d\eta^{\gamma}},
\end{equation}
The light-cone fraction variables $\mathcal{X}_g, \mathcal{X}_q, \mathcal{X}_q^{min}$ and $z_\gamma$ in Eq.~(\ref{pho},\ref{pho-f}) are 
defined as follows \cite{ja},
\begin{eqnarray} \label{x-3}
\mathcal{X}_g &=& \frac{1}{\mathcal{X}_q\, S} \left[{k_t^2\over z_\gamma} + \frac{|\q_T-\k_T|^2}{1-z_\gamma}\right],  \nonumber\\
z_\gamma&\equiv& \frac{k^-}{p^-} =\frac{k_T}{\mathcal{X}_q\, \sqrt{S}}\exp(\eta^{\gamma}),  \nonumber\\
 \mathcal{X}_q^{min}&=&\frac{k_T}{\sqrt{S}}\exp(\eta^{\gamma}).  \
\end{eqnarray}
Note that the
light-cone fraction variables defined above for the inclusive prompt photon cross-section Eq.~(\ref{x-3}) are different from the corresponding photon-jet and photon-hadron productions defined in Eqs.~(\ref{x-1},\ref{x-2}).

The main ingredient in the cross-section of semi-inclusive photon-jet (and photon-hadron) production in \eq{cs} and single inclusive prompt photon production in \eq{pho}  is the universal dipole forward scattering amplitude $N_F$ which incorporates small-x dynamics and can be calculated via the first-principle non-linear B-JIMWLK equations \cite{bk,jimwlk}. 
In the large $N_c$ limit, the coupled B-JIMWLK equations are simplified to the Balitsky-Kovchegov (BK) equation  \cite{bk}, a closed-form equation for the rapidity evolution of the dipole amplitude in which both linear radiative processes and non-linear recombination effects are systematically incorporated. The running-coupling BK (rcBK) equation has the following simple form \cite{nlo-bk1-2,rcbk}:
\begin{equation}
  \frac{\partial N_{F}(r,x)}{\partial\ln(x_0/x)}=\int d^2{\vec r_1}\
  K^{{\rm run}}(\r,\r_1,\r_2)
  \left[N_{F}(r_1,x)+ N_{F}(r_2,x)
- N_{F}(r,x)- N_{F}(r_1,x)\,N_{F}(r_2,x)\right]\,,
\label{bk1}
\end{equation}
where the evolution kernel $K^{{\rm run}}$ using Balitsky`s
prescription \cite{bb} for the running coupling is defined as,
\begin{equation}
  K^{{\rm run}}(\r,\r_1,\r_2)=\frac{N_c\,\alpha_s(r^2)}{2\pi^2}
  \left[\frac{1}{r_1^2}\left(\frac{\alpha_s(r_1^2)}{\alpha_s(r_2^2)}-1\right)+
    \frac{r^2}{r_1^2\,r_2^2}+\frac{1}{r_2^2}\left(\frac{\alpha_s(r_2^2)}{\alpha_s(r_1^2)}-1\right) \right],
\label{kbal}
\end{equation}
with $\r_2 \equiv \r-\r_1$. The only external input for the rcBK non-linear equation is the initial condition for the evolution which is taken to have the following form motivated by McLerran-Venugopalan (MV) model \cite{mv},  
  \begin{equation}
N(r,Y\!=\!0)=
1-\exp\left[-\frac{\left(r^2\,Q_{0s}^2\right)^{\gamma}}{4}\,
  \ln\left(\frac{1}{\Lambda\,r}+e\right)\right],
\label{mv}
\end{equation}
where the infrared scale is taken $\Lambda=0.241$ GeV and the onset of small-x evolution is assumed to be at
$x_0=0.01$ \cite{jav1}. The only free parameters in the above are $\gamma$ and the initial saturation scale $Q_{0s}$ probed by quarks with a notation $s=p$ and $s=A$ for a proton and a nucleus, respectively. The initial saturation scale of proton $Q_{0p}^2\simeq 0.168\,\text{GeV}^2$  with the corresponding $\gamma \simeq 1.119$ were extracted from a global fit to proton structure functions in DIS in the small-x region \cite{jav1} and single inclusive hadron data in p+p collisions at RHIC and the LHC  \cite{j1,me-jamal1,me-cgc,raj}.  Note that the current HERA data alone is not enough to uniquely fix the values of $Q_{0s}$ and $\gamma$ \cite{jav1}. The recent LHC data, however, seems to indicate that $\gamma>1$ is preferable \cite{j1}. 

Solving the small-x evolution equation in the presence of the impact-parameter is beyond the scope of weak-coupling perturbative framework and is currently an open problem in the  CGC approach \cite{bk-b}. Hence, in the rcBK evolution equation given in \eq{bk1}, the impact-parameter dependence of the collisions was ignored. The initial saturation scale $Q_{0s}$ should be then considered as an impact-parameter averaged value which effectively mimics the non-perturbative effects (such as impact-parameter dependence and possible fluctuations) and it is extracted from the experimental data.   For the minimum-bias collisions, one may assume that the initial saturation scale of a nucleus with atomic mass number A, scales linearly with $A^{1/3}$ \cite{mv}, namely we have $Q_{0A}^2=cA^{1/3}~Q_{0p}^2$ where the parameter $c$ is fixed from a fit to small -x data including DIS data on the nuclear target and duetron-nucleus and proton-nucleus data at RHIC and the LHC  \cite{me-jamal1,me-cgc,raj}. Note that a different $A$-dependence of the nuclear saturation scale with a pre-factor fitted to the HERA data, numerically leads to a very similar relation between the proton and nuclear saturation scale \cite{urs,me-cgc}.

The initial saturation of the nuclear target is generally defined as $Q^2_{0A}(x_0=0.01)=N_{part}^{Pb} Q^2_{0p}(x_0=0.01)$ where $N_{part}^{Pb}$ denotes the number of nucleon participants (or strictly speaking color charged probed) in the nuclear target side. The value of $N_{part}^{Pb}$ increases by decreasing the impact parameter of collisions from  medium impact-parameters (minimum bias) in the nucleus to very central impact parameters triggering rare high multiplicity events.

The initial saturation scale of heavy nuclei extracted from the current small-x data in {\it minimum-bias} heavy ion collisions, varies within\footnote{Throughout this paper, we use the convention to denote a range from $a$ to $b$ (or $[a,b]$) with $a\text{--}b$.}  $N_{part}^{Pb}\approx 4\text{--}7$ \cite{me-cgc}. Note that the recent LHC data for the nuclear modification of the charged hadrons $R_{pA}$ in minimum-bias collisions are in a good agreement with the solutions of the rcBK evolution equation with average initial saturation scale for nucleus with $N_{part}^{Pb}\approx 5$ \cite{me-cgc}. The high-multiplicity events can be described in the CGC approach by taking $N_{part}^{Pb}>5$. Such a large value for $N_{part}^{Pb}$ accounts the additional $A^{1/3}$ enhancement of the saturation scale due to probed hot-spot configurations of nucleus in central p+A collisions. For example, it was shown that a event with charged hadron multiplicity tracks within $90<N_{track}<110$ in p+Pb collisions at the LHC 5.02 TeV can be described by  $N_{part}^{Pb}\approx 14$ \cite{ridge4}.

In order to investigate the azimuthal angle correlations between the produced prompt photon and jet, we first calculate the coincidence probability. In the contrast to the di-hadron production, for the photon-jet (and photon-hadron) production we have freedom to select the trigger particle to be a produced prompt photon or a jet (or a hadron).  In a case that the trigger particle is a prompt photon,  the coincidence probability is defined as $CP(\Delta \phi)=N^{\text{photon-jet}}(\Delta \phi)/N^{\text{photon}}$, where $N^{\text{photon}}$  is the yield of single inclusive prompt photon, and $N^{\text{photon-jet}}(\Delta \phi)$ is the yield of photon-jet pair production including a associated quark jet with a transverse momentum $p^{asc}_{T}$ with a trigger (leading) prompt photon with transverse momentum $p^{trig}_{T}$ and the azimuthal angle between them  $\Delta \phi$ \cite{ja2}: 
\begin{eqnarray} 
CP(p_{T}^{\text{trig(photon)}}, \eta^\gamma, p^{\text{asc(jet)}}_{T},\eta^{jet},\Delta \phi)&=&\frac{2\pi \int_{p_T^{\text{trig}}}  dk_T k_T  \int_{p_T^{asc}} dq_T  q_T  \frac{d\sigma^{p+ A \rightarrow \gamma (k) + \text{jet}(q)+ X}} {d^2\b_T \, d^2\k_T\,d\eta^{\gamma}\, d^2\q_T \, d\eta^{jet}}    }  {{\int_{p_T^{\text{trig}}}  d^2\k_T   \frac{d\sigma^{p+A \rightarrow \gamma (k) + X}}{d^2\b_T d^2\k_T d\eta^{\gamma}}}} - C_{\text{ZYAM}}, 
 \label{cp1}\
\end{eqnarray}
where the integrals are performed within a given transverse momentum bin denoted by subscript $p^{trig}_{T}$ and  $p^{asc}_{T}$.  The yields in the above expression for photon-jet and single prompt photon production are given in Eqs.\,(\ref{qg-f},\ref{pho-f}). In the same fashion, one can define the coincidence probability with a jet-triggered as $N^{\text{photon-jet}}(\Delta \phi)/N^{\text{jet}}$ where $N^{\text{jet}}$ is the yield of single inclusive jet in p+A collisions calculated in the CGC approach \cite{hybrid}. However, the coincidence probability defined via selecting the trigger particle to be a jet is generally an order of magnitude smaller than the coincidence probability defined via the prompt photon-triggered in \eq{cp1}. This is simply due to the fact that the coupling $\alpha_{em}$ drops out in the ratio of \eq{cp1}.  Therefore, we only consider here coincidence probability defined via \eq{cp1}.  Following the experimental procedure for the zero-yield-at-minimum (ZYAM), we remove the constant background $C_{ZY AM}$ by shifting the minimum of the correlations at the zero axis.

In order to further examine the integrand of correlations defined via  \eq{cp1}, we compute the ratio of two-particle to single-particle cross-sections defined as follows, 
\begin{eqnarray} 
r_2(p_{T}^{\text{trig}}, \eta^\gamma, p_T^{\text{asc}},\eta^{jet},\Delta \phi)&=&\frac{2\pi  q_T^2  \frac{dN^{p+ A \rightarrow \gamma (k) + \text{jet}(q)+ X}} {d^2\k_T\,d\eta^{\gamma}\, d^2\q_T \, d\eta^{jet}}    }  {{ \frac{dN^{p+A \rightarrow \gamma (k) + X}}{d^2\k_T d\eta^{\gamma}}}}\Bigg{|}_{k_{T}=p_{T}^{\text{trig}},\, q_{T}=p_T^{\text{asc}}}.
 \label{r2}\
\end{eqnarray}
The ratio $r_2$  can be considered as a snap shot of the integrand in the coincidence probability defined in \eq{cp1} at fixed transverse momenta of the produced photon and jet. Following Refs.\,\cite{ja,ja2}, we also investigate the azimuthal correlations defined  in the following form, 
\begin{equation}\label{az}
C_2(k_T, \eta^\gamma, q_T,\eta^{jet},\Delta \phi)= {d\sigma^{q+A \rightarrow q(q)+\gamma(k)+ X}
\over d^2\b_T\, dk_T^2\,d\eta^{\gamma}\, dq_T^2\, d\eta^{jet}\, d\phi} /\int_0^\pi\,d\phi\,{d\sigma^{q(p)+A(p_A) \rightarrow q(q)+\gamma(k)+ X}
\over d^2\b_T\, dk_T^2\,d\eta^{\gamma}\, dq_T^2\, d\eta^{jet}\, d\phi} - C_{\text{ZYAM}}, 
\end{equation}
which has the meaning of the probability of the photon-jet pair production at a certain kinematics and angle  $\Delta \phi$, triggering the same production with the same kinematics (transverse momenta, rapidities and energies) integrated over all possible angles between photon-jet in $[0,2\pi]$. The correlation defined in \eq{az} may be more challenging to measure compared to the coincidence probability defined in \eq{cp1} due to the so-called underlying event dependence, but it is free from the extra integrals over transverse momenta, similar to \eq{r2}. Note that the normalization in the correlations defined in \eq{cp1}  and \eq{az} are different, and this facilitates to examine the effect of selecting different trigger for defining the correlations.

In this paper, we only consider observables which are defined as a ratio of cross-sections.  Note that we obtained the prompt photon cross-section \eq{pho} from photon-jet cross-section \eq{cs} by directly integrating over the jet 4-momenta without any approximation. We expect  that some of the theoretical uncertainties, such as sensitivity to $K$ factor which effectively incorporates the missing higher order corrections, will drop out in the azimuthal correlations defined in Eqs.\,(\ref{cp1}-\ref{az}). Our results are expected to be the same in the case that the projectile is a deuteron (rather than a proton) for deuteron-nucleus (d+A) collisions. This is mainly because the effect of the convolution with deuteron (or proton) structure functions drops out in the correlation functions defined via Eqs.\,(\ref{cp1}-\ref{az}).

We will use the NLO MSTW 2008 PDFs \cite{mstw} and the NLO KKP FFs \cite{kkp} for neutral pion. Following the conventional pQCD, we assume the factorization scale $Q$ in the PDF (and the FFs) to be equal to the transverse momentum of photon $k_T$ for inclusive prompt photon production. For the case of the photon-jet production the scale Q is taken to be largest transverse momenta in the system.

\section{Main results and discussions}
\begin{figure}[t]                                    
\includegraphics[width=10 cm] {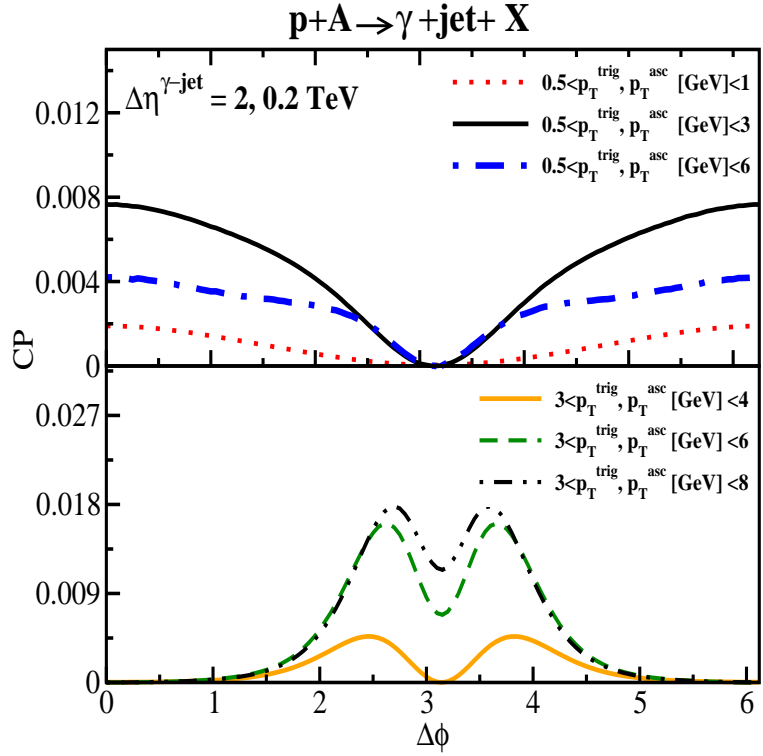}   
\caption{Transverse momentum dependence of photon-jet azimuthal correlations $CP$ defined  via the coincidence probability in \eq{cp1} as a function of angle between photon and jet $\Delta\phi$ at fixed rapidities of the produced photon  $\eta^{\gamma}=3$ and jet $\eta^{jet}=1$ (with $\Delta\eta^{\gamma-jet}=2$ ) at the RHIC energy 0.2 TeV.  Transverse momentum bins of the triggered photon $p_T^{trig}$ and associated  jet $p_T^{asc}$  are given in the plot. All results are obtained by the rcBK solutions with an initial saturation scale $Q^2_{0A}=15Q^2_{0p}$  (with $Q_{0p}^2=0.168\,\text{GeV}^2$) corresponding to a high-multiplicity event in p+p(A) collisions. }
\label{fig-cp-p}
\end{figure}

First, we study correlations of the produced photon and jet (with azimuthal angle between them $\Delta \phi$) by computing the coincidence probability defined in Eq.\,(\ref{cp1}).  
In \fig{fig-cp-p},  we show the coincidence probability for photon-jet production at a fixed rapidity of the produced jet and photon  $\eta^{jet}=1$ and  $\eta^{\gamma}=3$ (with $\Delta \eta^{\gamma-jet}=2$) at various transverse momenta of triggered photon $p_T^{trig}$ and associated jet $p_T^{asc}$ in p+p(A) collisions at RHIC energy 0.2 TeV. The results in \fig{fig-cp-p} are obtained by the solutions of the rcBK evolution equation (\ref{bk1}) with an initial saturation scale  $Q_{0A}^2=15~Q_{0p}^2$ (with $Q_{0p}^2=0.168\,\text{GeV}^2$) corresponding to a high-multiplicity event. It is seen that the near-side correlations at $\Delta \phi=0$  reach its maximum value in a kinematic window of $0.5<p_T^{trig}, p_T^{asc}[\text{GeV}]<3$, while at the other transverse momentum bins of $p_T^{trig}, p_T^{asc}$, the near-side correlations decrease and approach to zero for  $p_T^{trig}, p_T^{asc}> 3\,\text{GeV}$. At the away-side $\Delta \phi=\pi$, it is seen decorrelation of pairs at low transverse momenta of the produced photon and jet, and  the away-side correlation develops double peak structure at intermediate transverse momenta.

In order to understand the systematic features of our results, first note that the cross section of semi-inclusive photon-jet production in p+p(A) collisions given by \eq{hg-f} is proportional to
\begin{equation} \label{appx}
\frac{d\sigma^{p+ A \rightarrow \gamma (k) + \text{jet}(q)+ X}} {d^2\b_T \, d^2\k_T\,d\eta^{\gamma}\, d^2\q_T \, d\eta^{jet}}\propto |\q_T + \k_T|^2 N_F(b_T,  |\q_T + \k_T|, x_g).
\end{equation}
Therefore from the above it is seen that photon-jet production rate approaches to zero for:
\begin{equation} \label{ck}
 p_T^{total}=|\q_T + \k_T| \approx 0  \longrightarrow   \sigma^{p+ A \rightarrow \gamma (k) + \text{jet}(q)+ X} \approx 0. 
\end{equation}
This is simply because in order the higher Fock components of projectile hadron wavefunction to be resolved and a photon to be radiated,  the projectile quark should get a kick by interacting with small-x target via  exchanging transverse momentum $p_T^{total}$ to the target. Therefore, if the total transverse momentum exchanged to the target is zero $p_T^{total}=0$, the cross-section of photon-jet production goes to zero and off-shell photon remains as part of projectile hadron wavefunction. Note that in calculating the coincidence probability defined in Eq.\,(\ref{cp1}), because of convolution with parton distribution functions and integrals over the transverse momenta of the triggered and associated particles, the local minimum will not be zero but gets smeared out. Nevertheless, by subtracting of the zero-yield-at-minimum, one may reconstruct the minimum at zero. Obviously the condition given in \eq{ck} can be readily satisfied for away-side correlations at $\Delta \phi=\pi$ for $q_T\sim k_T$. On the other hand, the function $p_T^2N_F(p_T,x_g)$ in \eq{appx} and consequently the photon-jet cross-section has a maximum if the values of the total transfer transverse momentum $p_T^{\text{total}}$ and the saturation scale become similar, namely:
\begin{equation} \label{ck2}
 \text{If}\,\,\,\,\,\,p_T^{\text{total}}=|\q_T + \k_T|\approx Q_{s}\,\,\,\,\,\text{then}\,\,\,\,\,  \sigma^{p+ A \rightarrow \gamma (k) + \text{jet}(q)+ X} \longrightarrow \text{Maximum}.     
\end{equation}

\begin{figure}[t]                                    
\includegraphics[width=8 cm] {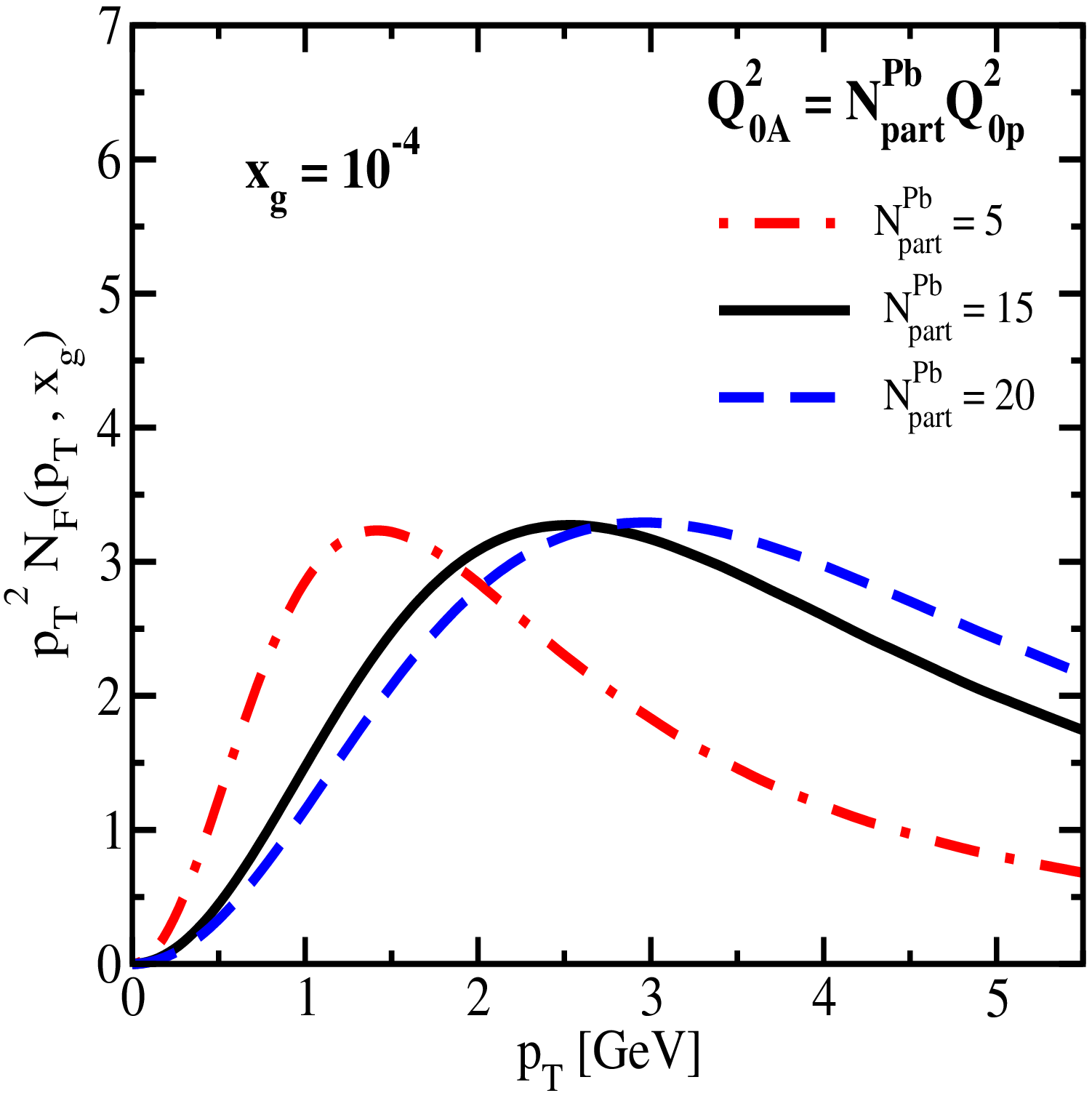}   
                                \includegraphics[width=8 cm] {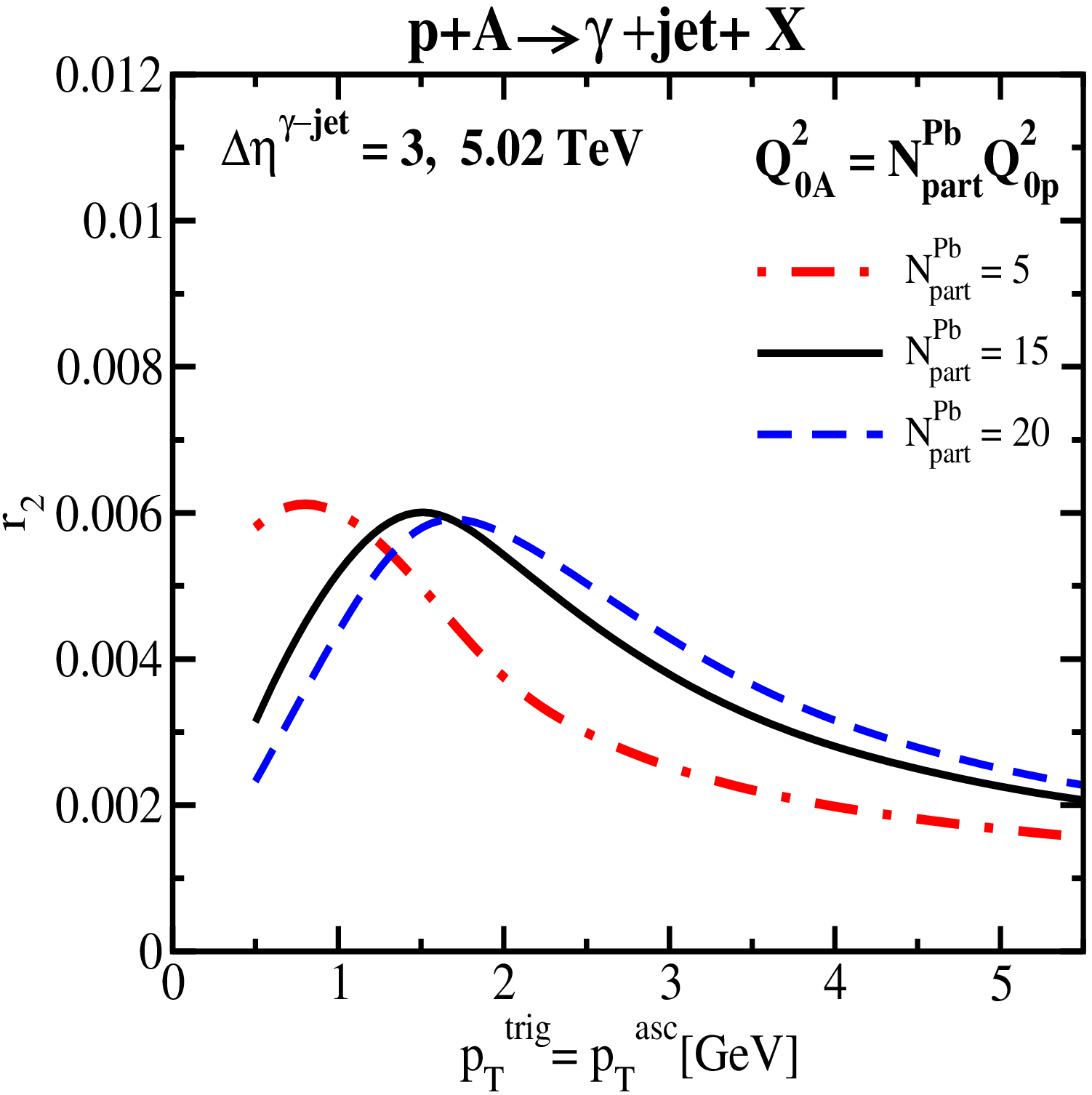}   
\caption{Left: transverse momentum dependence of function $p^2N_F(p_T,x_g)$ at a fixed $x_g=10^{-4}$ for different initial saturation scale of nuclei. Right: the ratio of photon-jet to single prompt photon cross-sections $r_2$ defined in \eq{r2} as a function of transverse momentum of triggered photon or associated jet for $p_T^{trig}=p_T^{asc}$  with rapidity interval  $\Delta\eta^{\gamma-jet}=3$ in p+A collisions at the LHC 5.02 TeV.  In both panels results are obtained by the rcBK solutions with different initial saturation scale $Q^2_{0A}=N_{part}^{Pb} Q^2_{0p}$ with $N_{part}^{Pb}=5,15,20$ and $Q_{0p}^2=0.168\,\text{GeV}^2$.   }
\label{fig-r2}
\end{figure}

In \fig{fig-r2} left panel we show the function $p_T^2N_F(p_T,x_g)$ obtained via the rcBK evolution equation, as a function of $p_T$ at a fixed $x_g$ for different initial-saturation scale $Q_{0A}$ defined via $Q^2_{0A}= N^{Pb}_{part} Q^2_{0p}$.  It is clearly seen that the function $p_T^2N_F(p_T,x_g)$ has a maximum about the saturation scale and the position of the maximum moves toward higher transverse momentum by increasing the initial saturation scale. Note that there is no unique convention for defining the actual saturation scale of a system at a given kinematic. The location of the peak of the phase space distribution of gluons in the transverse momentum space (shown in  \fig{fig-r2} left panel) determines the typical momentum of the gluons in the small-x hadron/nucleus wavefunction and is defined here as the saturation momentum.

Now returning to the systematic features of \fig{fig-cp-p}: for a kinematic bin with a similar transverse momentum of the photon (triggered particle) and jet (associated particle), $q_T\sim k_T$, the condition in \eq{ck2} can be satisfied at near-side $\Delta \phi=0$ when total transverse momentum $p_T^{total}$ becomes about the saturation scale of the system $Q_s$, while at the same time the condition in \eq{ck} can be satisfied for away-side at $\Delta\phi=\pi$. This leads to the enhancement or a maximum of the near-side correlations relative to the away-side correlations. For the other transverse momentum bins of prompt photon and jet that the total transfer momentum to the target is higher or lower than the saturation scale $p_T^{total}<Q_s$ or   
$p_T^{total}>Q_s$,  away from the condition given in \eq{ck2}, the near-side correlations at $\Delta\phi\approx 0$ diminishes as seen in \fig{fig-cp-p}. 
It is seen in \fig{fig-cp-p} that near-side correlations at $\Delta \phi\approx 0$  reach its maximum at the kinematics that the condition in \eq{ck2} is satisfied, namely when transverse momenta of the produced photon (triggered particle) and jet (associated particle) are within $0.5<p_T^{trig}, p_T^{asc}[\text{GeV}]<3$, and at other transverse momentum bins (shown in \fig{fig-cp-p}), the near-side correlations is smaller. 

\begin{figure}[t]                                    
\includegraphics[width=8 cm] {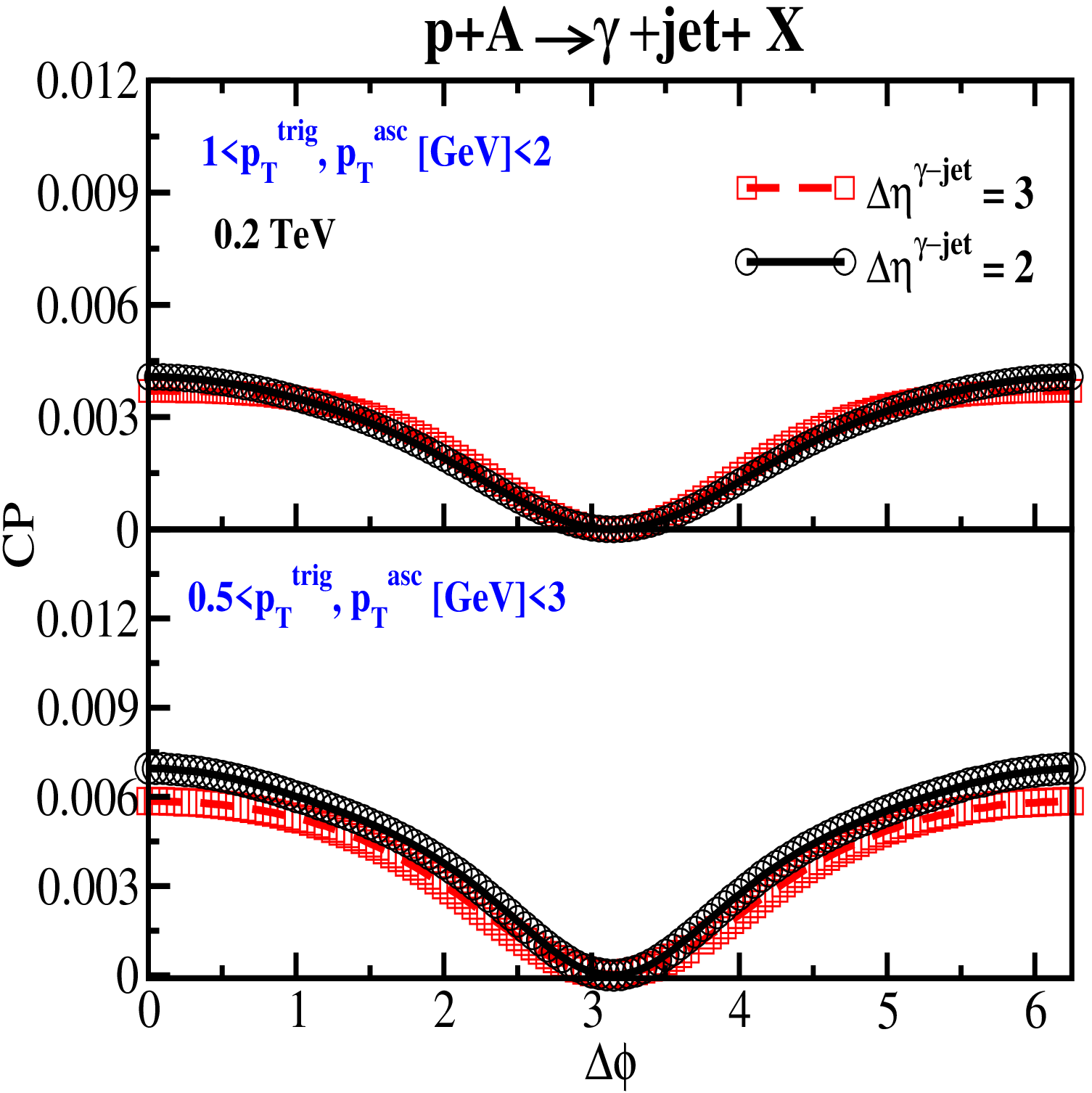}   
                \includegraphics[width=8 cm] {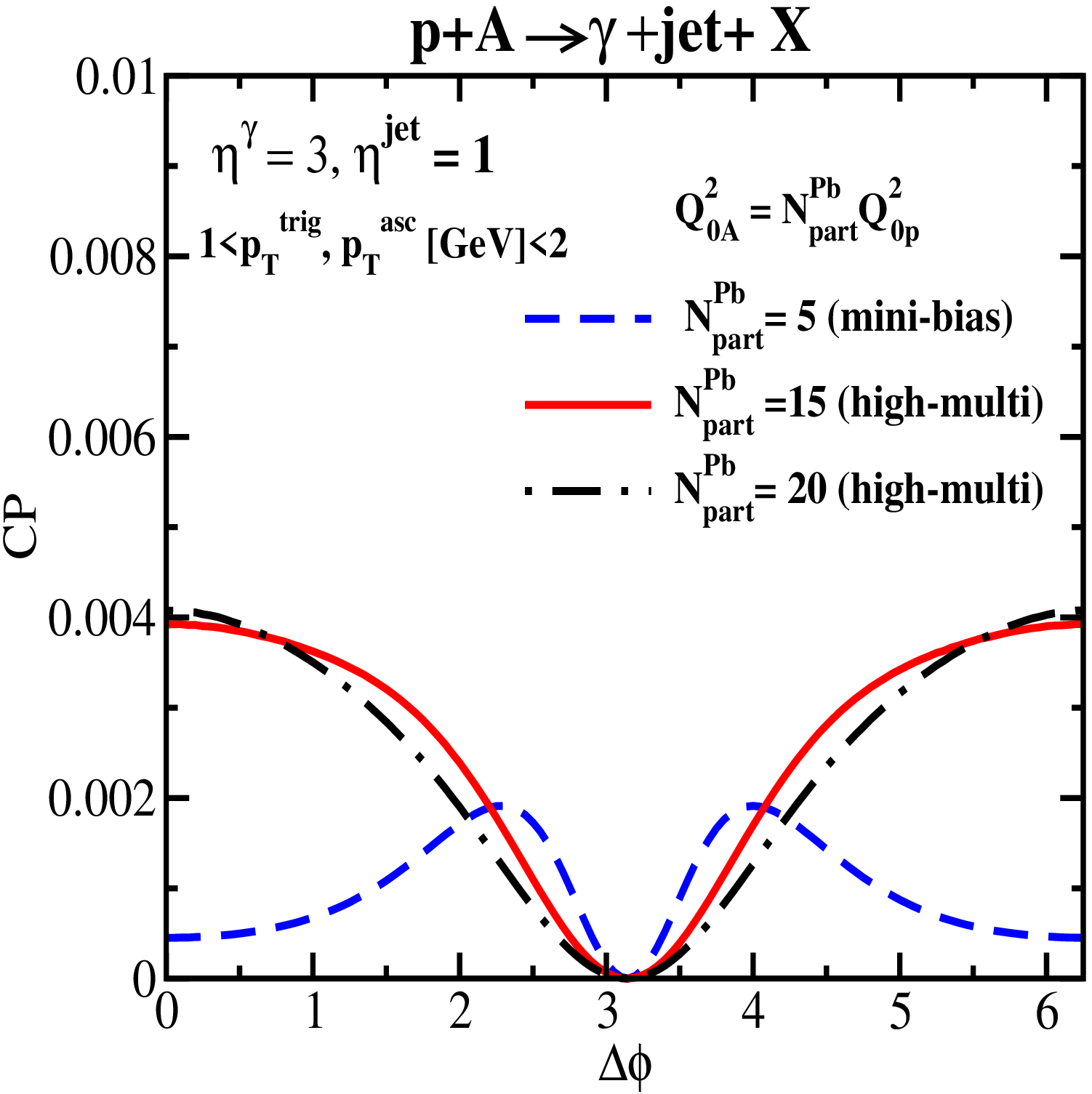}                                     
\caption{ Left: $\gamma$-jet correlations $CP(\Delta \phi)$ defined via the coincidence probability in \eq{cp1} within two transverse momentum bins at two rapidities intervals  $\Delta\eta^{\gamma-jet}=2, 3$  in p+A collisions at RHIC.   
Right: initial-saturation scale $Q_{0A}$ dependence of $CP(\Delta \phi)$ at a fixed transverse momentum bin and rapidity interval $\Delta\eta^{\gamma-jet}=2$ (with  $\eta^{jet}=1$) in p+A collisions at RHIC. In right panel,  the results obtained with the rcBK solutions with different initial saturation scale $Q^2_{0A}=N_{part}^{Pb} Q^2_{0p}$ where $N_{part}^{Pb}=5,15,20$ is the number of nucleon participants on the lead (or gold) side. In left panel, all results are obtained with a fixed $N_{part}^{Pb}=15$. The initial saturation scale of proton is taken $Q_{0p}^2=0.168~\text{GeV}^2$. }
\label{fig-cp-eta}
\end{figure}

For higher transverse momenta, the saturation physics is negligible and the back-to-back correlation is restored\footnote{
At high transverse momentum the photon-jet cross-section matches to the pQCD expressions \cite{pho-cgc2}, including the standard back-to-back correlations. At low transverse momentum at the small-x region, the existence of an extra scale into the system, namely the saturation scale, unbalances the back-to-back correlations, leading to the decorrelation or the suppression of back-to-back correlations. } and consequently there would be a local maximum at away-side while at the same time, the condition  in \eq{ck} results to a local minimum at away-side. The interference of a local-maximum and a local minimum at $\Delta \phi=\pi$  leads to a double-peak structure, as it is seen  in \fig{fig-cp-p}.  Note that at the kinematics that the condition in \eq{ck} is not satisfied,  the double peak structure at away-side is fused to a single  away-side peak (not shown here). At low transverse momenta in high multiplicity events, the saturation scale is large and the existence of a large extra scale in the system unbalances the back-to-back correlations.

In \fig{fig-r2} (right), we show the ratio of photon-jet  to single prompt photon production $r_2$ defined in \eq{r2} at near-side $\Delta \phi=0$ at forward rapidity $\eta^{\gamma}=4$ and  $\eta^{jet}=1$ (with $\Delta \eta^{\gamma-jet}=3$) as a function of transverse momentum of triggered photon (or associated jet) for $p_T^{trig}=p_T^{asc}$ at  the LHC energy 5.02 TeV. The effect of events with different centrality or multiplicity  are simulated by employing the solutions of the rcBK evolution equation with different initial saturation scales for the target  $Q^2_{0A}= N^{Pb}_{part} Q^2_{0p}$ with $ N^{Pb}_{part}=1, 5,15$ and $Q_{0p}^2=0.168~\text{GeV}^2$. In \fig{fig-r2} (left), we show the dipole unintegrated gluon density distribution as a function of transverse momentum at a fixed $x_g=10^{-4}$ (corresponding to the LHC energy). Comparing right and left panels in \fig{fig-r2}, it is seen that the relative variation of dipole unintegrated gluon density profile with the saturation scale (and density) leads to a similar effect for the function $r_2$ defined in \eq{r2} and consequently for the integrand of correlation function $CP$ defined in \eq{cp1}. It is also seen that the existence of a maximum for the function $p_T^2N_F(p_T,x_g)$ (in left panel) at a transverse momentum $p_T^{\max}$ leads to a maximum for the ratio $r_2$ at $\Delta \phi=0$  at about $p_T^{trig}\approx p_T^{\max}/2$ (in right panel), consistent with Eqs.\,(\ref{appx},\ref{ck2}). 

\begin{figure}[t]                                    
\includegraphics[width=8 cm] {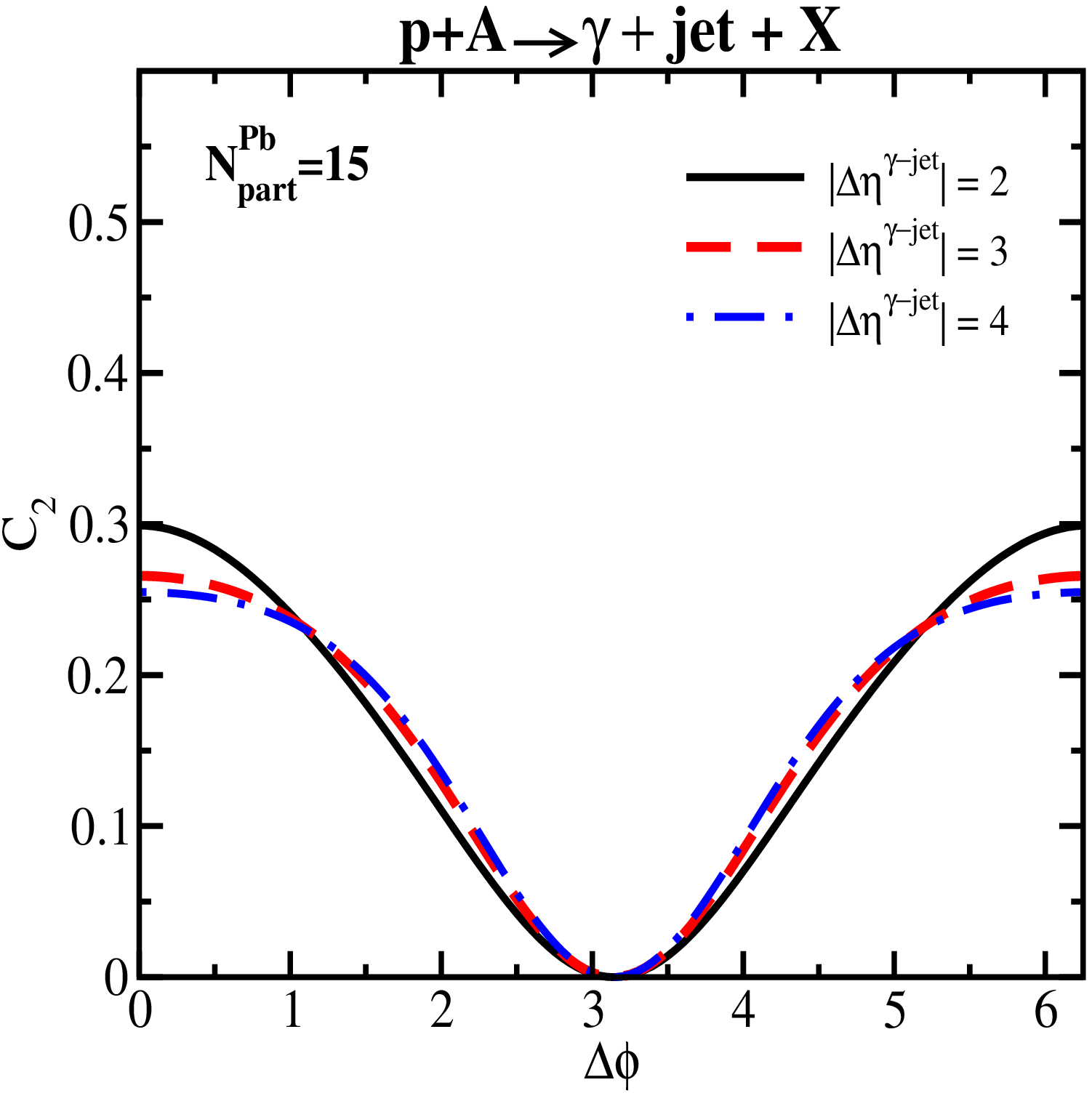}   
                 \includegraphics[width=8 cm] {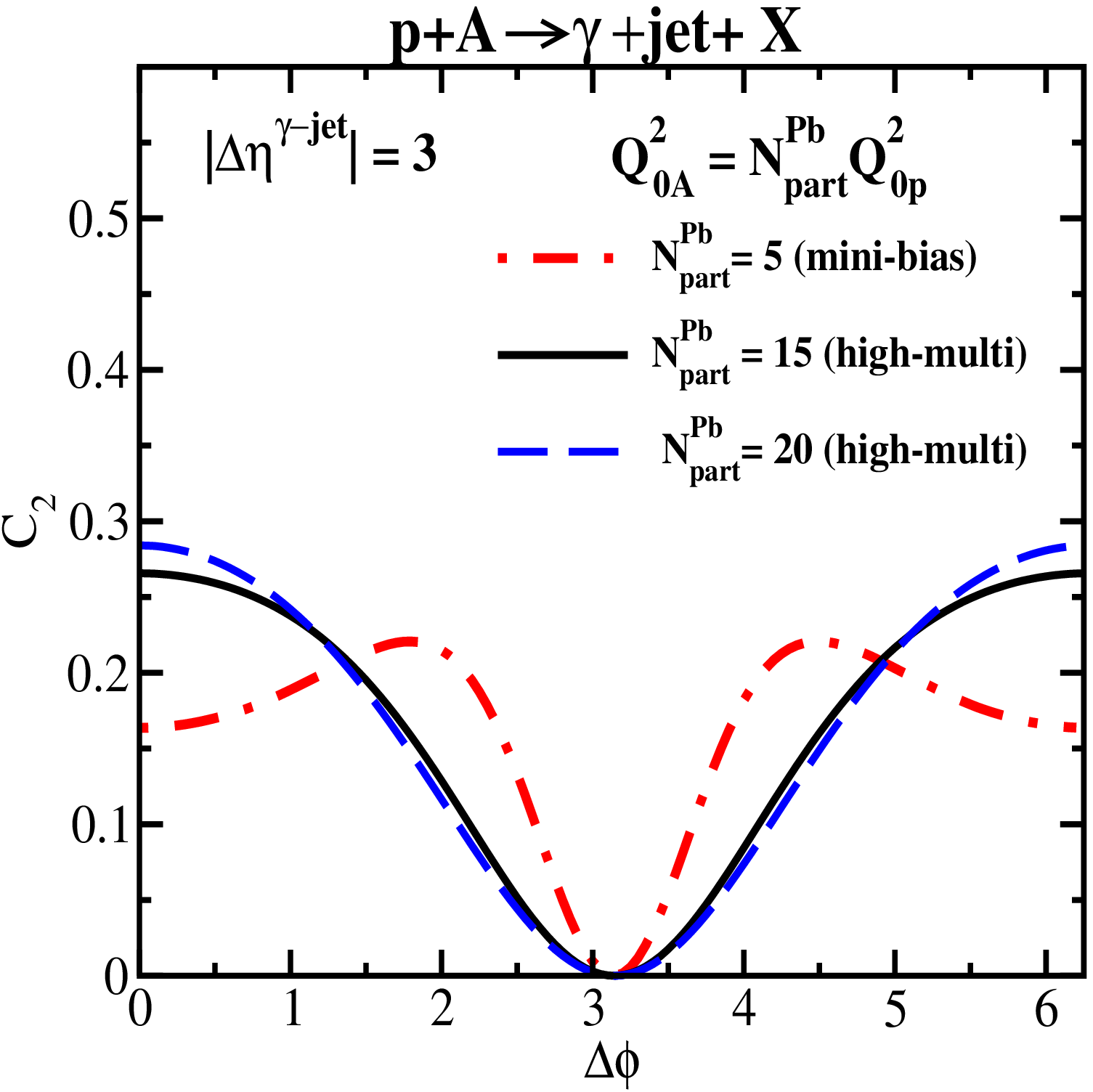}     
\includegraphics[width=8 cm] {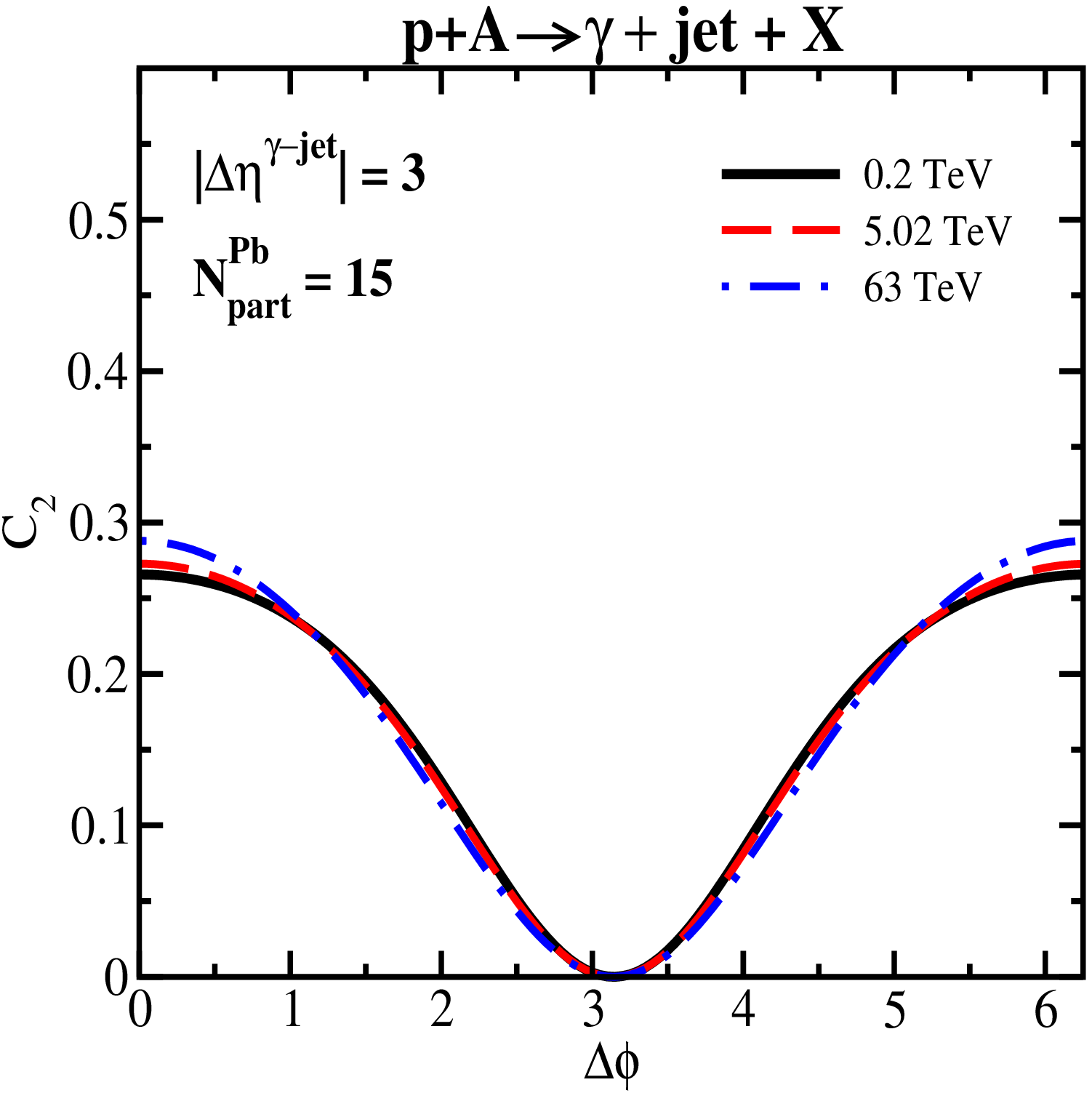}                                
\caption{Top left: rapidity-interval $|\Delta\eta^{\gamma-jet}|$ dependence of photon-jet azimuthal correlation $C_2(\Delta \phi)$ defined in \eq{az} as a function of angle between the produced photon and jet $\Delta \phi$ in p+A collisions at 0.2 TeV.  Top right: initial-saturation scale $Q_{0A}$ dependence (or multiplicity dependence) of  $C_2(\Delta \phi)$ for a fixed rapidity interval $|\Delta \eta^{\gamma-jet}|=3$ at RHIC obtained by the rcBK solutions with different initial saturation scale $Q^2_{0A}=N_{part}^{Pb} Q^2_{0p}$ with $N_{part}^{Pb}=5,15,20$ and $Q_{0p}^2=0.168~\text{GeV}^2$. Lower:  energy dependence of  $C_2(\Delta \phi)$ at a fixed rapidity interval $|\Delta \eta^{\gamma-jet}|=3$ at different energies 0.2, 5.02 and 63 TeV in p+A collisions. In top left and lower panels all results are obtained with a fixed $N_{part}^{Pb}=15$. In all panels,  we take transverse momenta of the produced photon and jet at $k_T=q_T=1$ GeV and rapidity of the produced jet at $\eta^{\text{jet}}=1$.  }
\label{fig-c2-eta}
\end{figure}

In \fig{fig-cp-eta} (left), we show the rapidity-interval $\Delta \eta^{\gamma-jet}$ dependence of photon-jet correlations $CP$ defined via  \eq{cp1} 
as a function of angle between photon and jet $\Delta \phi$, within two transverse momenta bins in which the near-side correlations is maximum, namely for $0.5<p_T^{trig}, p_T^{asc}[\text{GeV}]<3$ and $1<p_T^{trig}, p_T^{asc}[\text{GeV}]<2$ in p+A collisions at RHIC. All curves in \fig{fig-cp-eta} (left) are obtained at a fixed rapidity of jet $ \eta^{jet}=1$. The rapidity of photon is given by $\eta^{\gamma}=\Delta \eta^{\gamma-jet}+\eta^{jet}$.  In left panel, the initial saturation scale of the nuclear target is taken $Q^2_{0A}= 15\text{--} 20 Q^2_{0p}$ corresponding to high-multiplicity events. It is remarkably seen that the near-side and away-side correlations practically do not change within $\Delta \eta^{\gamma-jet} \approx 2\text{--} 4$ indicating that the photon-jet correlations in high-multiplicity events are characteristically long-range in rapidity interval (see also \fig{fig-c2-eta}). In  \fig{fig-cp-eta} (right), we show the multiplicity (or density) dependence of the photon-jet correlations $CP$ for different values of $N^{Pb}_{part}=5, 15, 20$. Note that the initial saturation scale of the nuclear target  with $N^{Pb}_{part}>5$, corresponds to events with higher charged hadron multiplicity than in minimum bias collisions. We recall that the di-hadron ridge type structure is well pronounced in multiplicity events corresponding to $N^{Pb}_{part}=15$  \cite{ridge4}. It is seen from \fig{fig-cp-eta} right panel that the near-side correlations at $\Delta \phi=0$ is sensitive to the saturation scale (and multiplicity event selection): it dramatically enhances and reaches its maximum in high multiplicity collisions for $N^{Pb}_{part}\ge15$, and becomes practically negligible for about $N^{Pb}_{part}\le 5$ corresponding to minimum-bias collisions or  peripheral p+A collisions.

\begin{figure}[t]       
                               \includegraphics[width=8 cm] {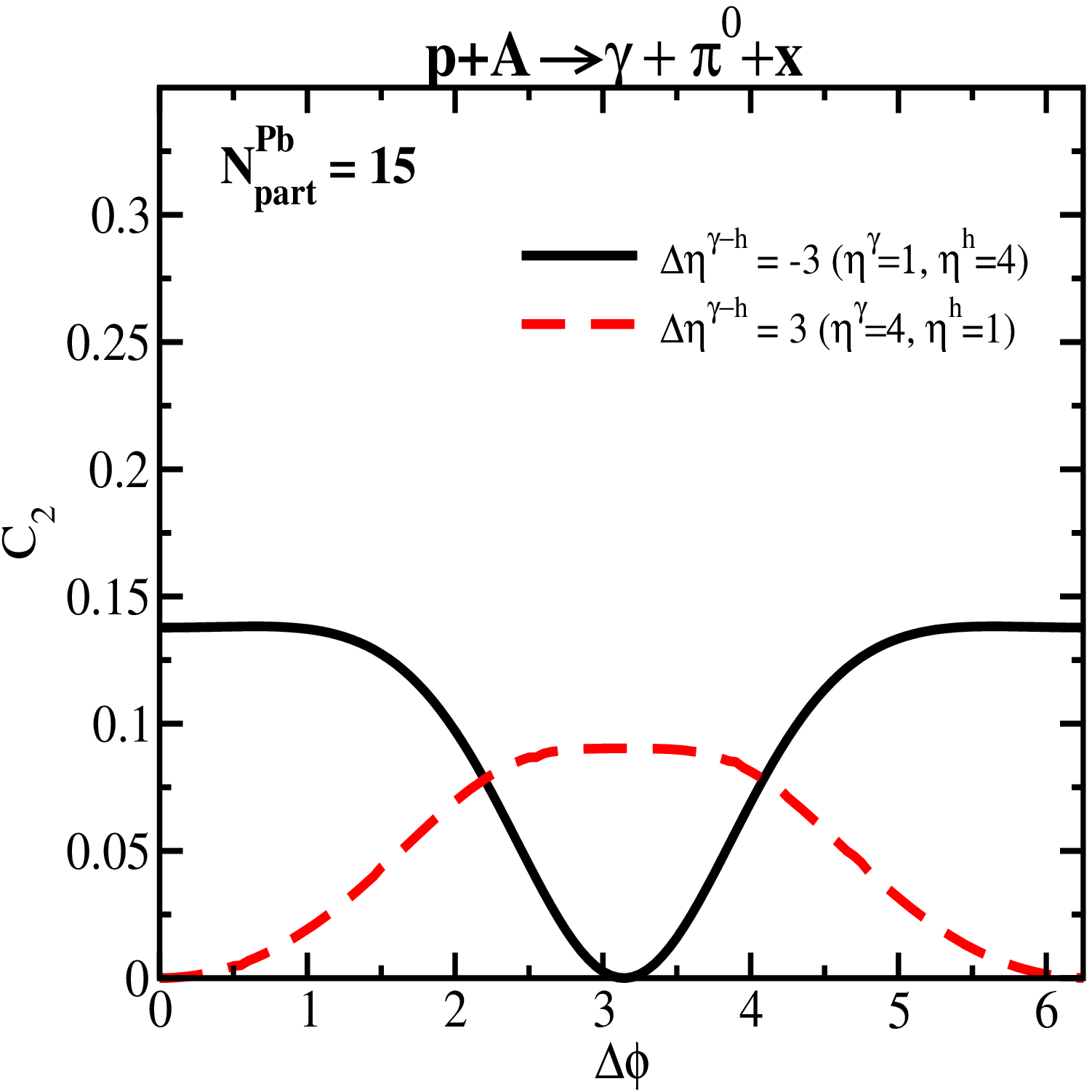}    
\includegraphics[width=8 cm] {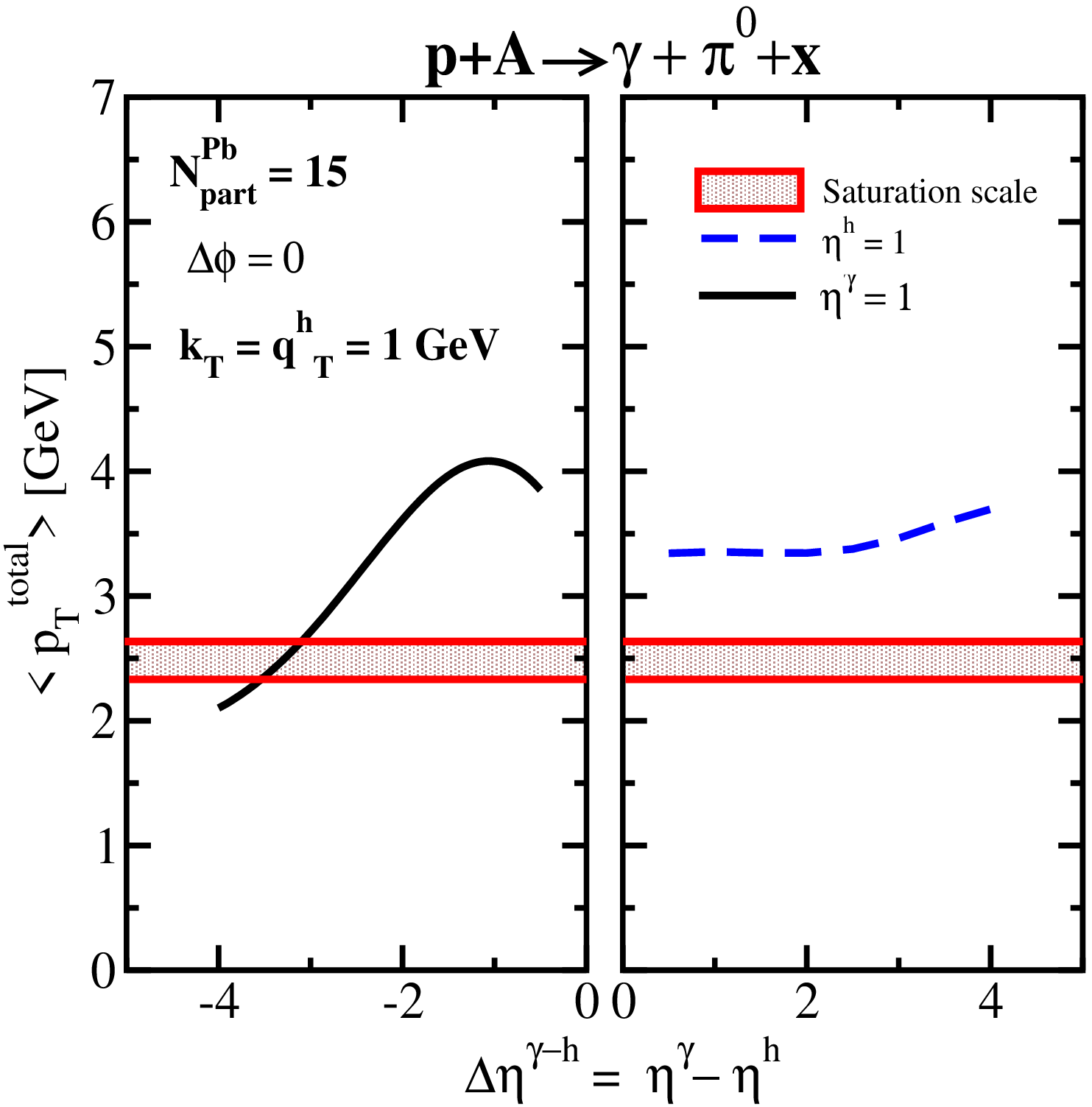}                                    
\caption{ Left: azimuthal photon-hadron ($\gamma-\pi^0$) correlation $C_2$  as a function of angle between the produced hadron and prompt photon $\Delta \phi$ at two rapidities intervals of $\Delta \eta^{\gamma-h}=-3$  (corresponding to $ \eta^{\gamma}=1$ and $\eta^{h}=4$) and $\Delta \eta^{\gamma-h}=3$  (corresponding to $ \eta^{\gamma}=4$ and $\eta^{h}=1$) in p+A collision at RHIC. Right: average total transverse momentum defined by $\langle p_T^{total}=|\k_T+\q_T/z_h|\rangle$
 at $\Delta \phi=0$ as a function of  $\Delta \eta^{\gamma-h}= \eta^{\gamma}- \eta^{h}$ for two cases: once with a fixed rapidity of photon $\eta^{\gamma}=1$ and in the other case with a fixed rapidity of hadron $\eta^{h}=1$.  
In both panels, the results are obtained by the rcBK evolution solutions assuming the initial-saturation scale of the target to be $Q^2_{0A}=15Q^2_{0p}$  corresponding to a high-multiplicity event with the number of nucleon participants on the lead side of $N_{part}^{Pb}=15$. In both panels, we take  transverse momenta of the produced photon and hadron at $k_T=q_T^h=1$ GeV. }
\label{fig-c2-h2}
\end{figure}
Next, we consider the photon-jet correlation $C_2$ defined via \eq{az}. We recall that the correlation functions $C_2$ and $CP$ defined in Eqs.\,(\ref{cp1},\ref{az}) are different in the definition of the trigger particle (and the normalization). Moreover, in contrast to the CP function, the correlation function $C_2$ is defined at fixed transverse momentum of photon and jet without invoking any extra integration over the transverse momenta of the produced photon and jet. In \fig{fig-c2-eta}, we show the rapidity interval  $\Delta \eta^{\gamma-jet}$ (top-left panel) and the initial saturation scale dependence (top-right panel) of the photon-jet correlation $C_2$  with $k_T=q_T=1$ GeV in p+A collisions at 
RHIC energy 0.2 TeV. Similar to the correlation defined via the coincidence probability $CP$ shown in \fig{fig-cp-eta}, it is seen in \fig{fig-c2-eta} (top panel) that the photon-jet correlation $C_2$ at near-side $\Delta\phi=0$ increases with the initial saturation scale (or multiplicity) and becomes  independent of rapidity interval at high-multiplicity events (corresponding to about $N_{part}^{Pb}\ge15$).

In \fig{fig-c2-eta} lower panel, we show the energy dependence of $C_2$ as a function of $\Delta\phi$ at a fixed rapidity separation $\Delta \eta^{\gamma-jet}=3$ at RHIC energy 0.2 TeV, the LHC energy 5.02 TeV and the Future Circular Collider (FCC) energy 63 TeV  \cite{fcc} in p+A collisions. In both top-left and lower panels, we employed the photon-jet cross-section supplemented with the rcBK solution with an initial-saturation scale corresponding to $N_{part}^{Pb}=15$. Note that at the same value of $N_{part}^{Pb}=15$, we observe ridge like structure for the photon-jet correlations. It is remarkable that while the variable $x_g$  which appears into the dipole amplitude and defined in \eq{x-1}, changes from about $10^{-3}$ at RHIC 0.2 TeV to $10^{-6}$ at the FCC 63 TeV, there is little energy dependence in the photon-jet correlation $C_2$ at high-multiplicity events.

In a similar fashion, one can also study photon-hadron correlation using \eq{az} supplemented by \eq{hg-f}. First note that the correlation of photon-jet $C_2$ defined via Eqs.\,(\ref{qg-f},\ref{az}) are symmetric  with respect to swapping the produced photon and jet in rapidity and momentum space, namely by the following replacement, 
\begin{equation}
C_2\left(k_T, \eta^\gamma, q_T,\eta^{jet},\Delta \phi\right)=  C_2\left(k_T \to q_T, \eta^\gamma\to\eta^{jet} , q_T\to k_T,\eta^{jet}\to\eta^\gamma ,\Delta \phi\right).
\end{equation}
In a sharp contrast, because of the convolution with quark-hadron fragmentation functions and extra integral over the fragmentation parameter $z_h$ in the cross-section \eq{hg-f}, the photon-hadron correlation $C_2$ defined via  Eqs.\,(\ref{hg-f},\ref{az})  is not symmetric:  
\begin{equation}
C_2\left(k_T, \eta^\gamma, q_T^{h},\eta^{h},\Delta \phi\right)\ne C_2\left(k_T \to q_T^h, \eta^\gamma\to\eta^{h} , q_T^h\to k_T,\eta^{h}\to\eta^\gamma ,\Delta \phi\right).
\end{equation}
 This leads to nontrivial effects for photon-hadron correlations as compared to the symmetric productions such as di-hadron or prompt di-photon. Note also that the cross-section of photon-jet production and the correlation defined via coincidence probability in \eq{cp1} are not symmetric with respect to the replacement of $(\eta^\gamma\longleftrightarrow\eta^{jet}, q_T\longleftrightarrow k_T)$. 

\begin{figure}[t]       
                               \includegraphics[width=8 cm] {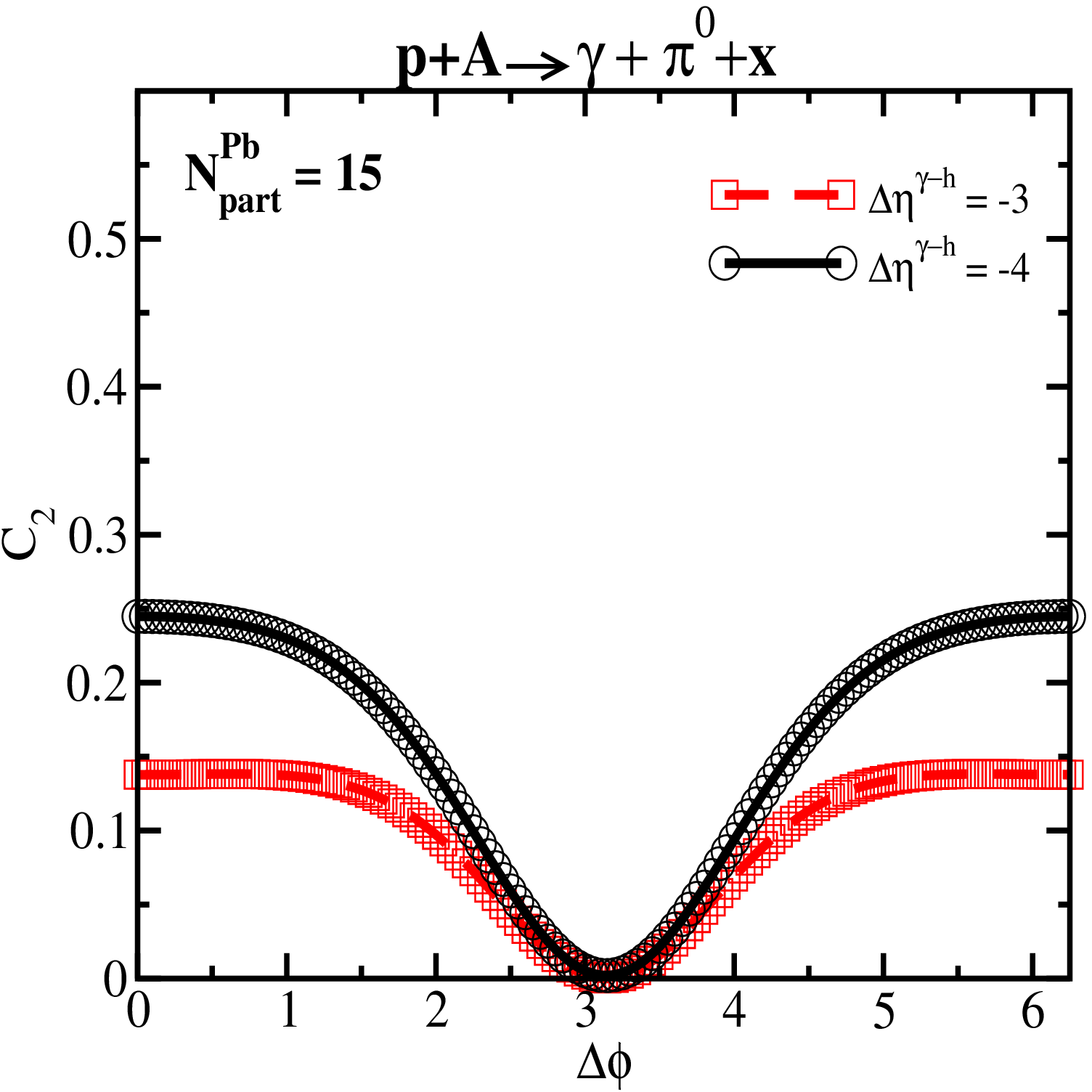}       
                                       \includegraphics[width=8 cm] {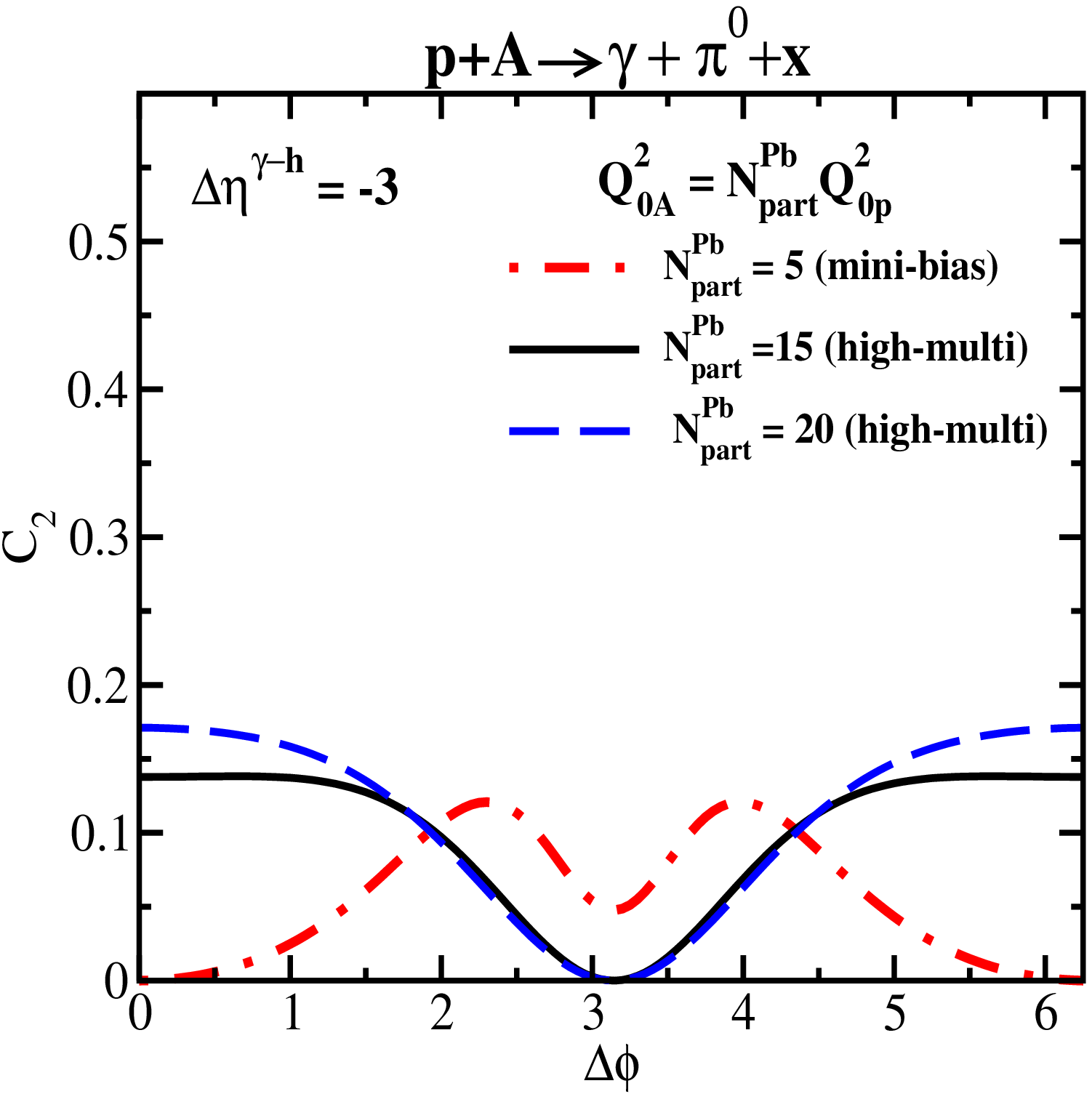}                     
\caption{ Left: rapidity-interval dependence of the photon-hadron azimuthal correlation $C_2(\Delta \phi)$  as a function of angle between the produced photon and jet $\Delta \phi$ in p+A collisions at 0.2 TeV.  Right: initial-saturation scale $Q_{0A}$ dependence (or multiplicity dependence) of   photon-hadron $C_2(\Delta \phi)$ at a fixed rapidity interval $\Delta \eta^{\gamma-jet}=-3$ in p+A collisions at 0.2 TeV, obtained by the rcBK solutions with different initial saturation scale $Q^2_{0A}=N_{part}^{Pb}Q^2_{0p}$ with $N_{part}^{Pb}=5,15,20$ and $Q_{0p}^2=0.168~\text{GeV}^2$.  In left panel the results are obtained with a fixed $N_{part}^{Pb}=15$.  In both panels, we take transverse momenta of the produced photon and hadron at $k_T=q_T^h=1$ GeV and rapidity of hadron at $\eta^{h}=1$.   }
\label{fig-c2-h1}
\end{figure}

In \fig{fig-c2-h2} left panel, we show the photon-$\pi^0$ correlation $C_2$ as a function of $\Delta \phi$ at fixed transverse momenta of photon and hadron $k_T=q_T^h=1$ GeV  in two rapidities interval cases: $\Delta \eta^{\gamma-h}=-3$ ($\eta^{\gamma}=1, \eta^h=4$) and  $\Delta \eta^{\gamma-h}=3$ ($\eta^{\gamma}=4, \eta^h=1$) where  we defined $\Delta \eta^{\gamma-h}=\eta^{\gamma}-\eta^{h}$.  In both cases we used the rcBK solutions for the dipole unintegrated gluon density with $N^{Pb}_{part}=15$. The correlation in these two (seemingly symmetric) bins are drastically different: the near-side correlations only exist for the case of $\Delta \eta^{\gamma-h}=-3$  and it disappears in the case of $\Delta \eta^{\gamma-h}=3$ while the away-side back-to-back correlation is only significant in the case of $\Delta \eta^{\gamma-h}=3$.  
In order to understand the origin of these effects, in \fig{fig-c2-h2} right panel, we show the average total transfer (transverse) momentum $\langle p_T^{total}=|\k_T+\q_T/z_h|\rangle$ as a function of  $\Delta \eta^{\gamma-h}$ at $\Delta \phi=0$.  Similar to left panel, the results in the right panel are obtained with $k_T=q_T^h=1$ GeV
for two cases of  $\eta^h=1$ (and $\Delta \eta^{\gamma-h}>0$) and 
$\eta^\gamma=1$ (and $\Delta \eta^{\gamma-h}<0$). We also show in the plot, the saturation scale as a function of $\Delta \eta^{\gamma-h}$. 
Note that $\langle z_h\rangle$ becomes large and approaches to one at very forward rapidity of hadron $\eta^h$. Therefore, the average total transverse momentum $\langle p_T^{total}\rangle$ approaches to its minimum value for $\Delta \eta^{\gamma-h}<0$. In the contrast if we increase the rapidity of the photon (for $\Delta \eta^{\gamma-h}>0$) but keep the rapidity of hadron and transverse momenta of photon and hadron fixed, then the average hadron fragmentation parameter $\langle z_h\rangle$ and consequently $\langle p_T^{total}\rangle$  do not change that much. These effects are seen in \fig{fig-c2-h2} right panel. It is seen that for the case of $\Delta \eta^{\gamma-h}<0$ the average total transverse momentum $\langle p_T^{total} \rangle$ decreases with increasing the rapidity of hadron and at rapidity interval about $\Delta \eta^{\gamma-h}=-3$ become equal to the saturation scale. At this kinematics, the condition in \eq{ck2} will be satisfied and as a result the near-side correlation get enhanced. This is the main reason behind the fact that the ridge-like structure only exist for $\Delta \eta^{\gamma-h}<0$ (shown in left panel in \fig{fig-c2-h2}). For a similar reason, since average total transverse momentum $\langle p_T^{total} \rangle$ is bigger than the saturation scale for the case of  $\Delta \eta^{\gamma-h}>0$, there is no a ridge-type structure at $\Delta \phi=0$ and the away-side back-to-back correlations are restored and consequently away-side correlations get enhanced at $\Delta \phi=\pi$ leading to a peak structure.

In \fig{fig-c2-h1}, we show the azimuthal correlation $C_2$ defined via \eq{az} for $\gamma-\pi^0$ as a function of angle between prompt photon and hadron $\Delta\phi$ at different rapidity interval $\Delta \eta^{\gamma-jet}$ (left) for a fixed initial saturation scale corresponding to $N^{Pb}_{part}=15$, and for different initial saturation scale (right) at a fixed rapidity interval $\Delta \eta^{\gamma-jet}=-3$ in p+A collisions at the RHIC energy of 0.2 TeV. In both panels in \fig{fig-c2-h1}, the transverse momenta of photon and hadron are taken  to be $k_T=q_T^h=1$ GeV and rapidity of hadron is fixed at $\eta^{h}=1$.  It is seen from \fig{fig-c2-h1} that in contrast to photon-jet correlations shown in \fig{fig-c2-eta}, the variation of the near-side photon-hadron correlations with rapidity interval is larger. It is also seen in \fig{fig-c2-h1} (right panel) that the photon-hadron near-side correlations totally wash away in minimum-bias collisions with $N^{Pb}_{part}\approx 5$, and only exist at high multiplicity collisions with $N^{Pb}_{part}\gg 5$.

Finally note that Ref.\,\cite{ridge4} required both the projectile proton and the target to be in the saturation regime in order to describe the ridge effect in p+Pb collisions.  The existence of two saturation scales in the problem for the projectile proton and the target  both in p+p and p+Pb collisions, is essential  for Ref.\,\cite{ridge4} in order to find a good fit to the observed number of charged hadron tracks in which the ridge appears. The origin of such rare high multiplicity events in asymmetric collisions such as p+Pb collisions is not yet well understood. In our hybrid approach on the other hand, we only focused at forward rapidity production, assuming that the projectile proton is in dilute regime and can be consequently treated in standard parton model approach, while the target is in gluon saturation regime and can be treated in the CGC approach. Therefore, we have here only one saturation scale in our system. Nevertheless, remarkably we found that a ridge-like correlation exists for photon-jet pair production in high-multiplicity collisions at large $N^{pb}_{part}$. 

\section{Conclusion}
In this paper, we investigated the existence of ridge-type correlations for photon-jet and photon-hadron production within the CGC approach in high-energy  p+p(A) collisions at RHIC and the LHC. Our formulation is only valid for the case of dilute-dense scatterings at very forward rapidities where the target can be treated in the small-x saturation regime while the projectile proton is in dilute regime and mainly dominated by quarks and antiquarks.

We found that a ridge-type structure exists for pairs of prompt photon and jet production in high-multiplicity p+A collisions at both RHIC and the LHC. 
We showed that the collimation of long-range in the relative rapidity $\Delta \eta^{\gamma-jet}$ of pairs of prompt photon and jet at the near-side $\Delta\phi\approx 0$, the ridge effect, strongly depends on the saturation dynamics and is directly related to the existence of a maximal occupancy (a peak) in the phase space distribution of gluons in the target wavefunction (or gluon saturation). The photon-jet collimation at near-side is maximum  at kinematics that the value of typical total transfer momentum to the target is similar to the value of the saturation scale in the system $p_T^{\text{total}}\approx Q_s$, see \eq{ck2}. We showed that the ridge effect only appears in a small kinematic window at low transverse momenta of the produced jet and photon about $1\text{--} 3$ GeV in high multiplicity events with  $N^{Pb}_{part}>5$ at RHIC and the LHC energies. The effect should be clearly visible for $N^{Pb}_{part}\approx 15\text{--} 20$, see Figs.\,(\ref{fig-cp-eta},\ref{fig-c2-eta},\ref{fig-c2-h1}).  
Within such a kinematic window, the energy dependence of the photon-jet ridge is very little, see \fig{fig-c2-eta}. We also showed that the ridge extends within few rapidity intervals between photon and jet, namely for $\Delta \eta^{\gamma-jet}\approx 2\text{--} 4$.  These features are strikingly similar to the observed ridge effect for di-hadron correlations at RHIC and the LHC.

We also studied the near-side photon-hadron correlations, and showed that the hadronization of the jet has non-trivial effects on the correlation. The photon-hadron correlations defined via Eqs.\,(\ref{cp1},\ref{az}) is not symmetric with respect to replacing the rapidities of the produced prompt photon and hadron ($\eta^h \longleftrightarrow\eta^{\gamma}$) when transverse momentum of photon and hadron is equal $k_T=q_T^h$. This is in a contrast to symmetric cases like di-hadron and di-photon productions.  We showed that photon-hadron near-side collimation at $\Delta \phi=0$ (and the ridge effect) only exists for the cases of $\Delta \eta^{\gamma-h}=\eta^\gamma-\eta^h<0$ in high-multiplicity collisions with a large $N^{Pb}_{part}\gg 5$. We showed that this effect can be traced back to the rapidity dependence of average total transfer momentum $\langle p_T^{total} \rangle$ satisfying the condition in \eq{ck2}, see \fig{fig-c2-h2}.

The prompt photon and jet pair correlations at forward rapidities in the away-side region exhibit decorrelation at low transverse momenta, and 
double (or single) peak structure at high transverse momenta, see \fig{fig-cp-p}.  The existence of a double-peak structure at away-side seems to be a universal feature of Electromagnetic probes, and was also reported for photon-hadron \cite{ja2}, dilepton-hadron \cite{dy-ana}, and di-photon \cite{di-photon} correlations in high-energy p+Pb collisions.

Although di-hadron ridge was observed  in p+p and p(d)+A collisions \cite{exp-pp,exp-pa1,exp-pa2,exp-pa3,exp-pa4,exp-pa5,exp-pa6} both at the LHC and RHIC, the photon-jet and photon-hadron correlations have not yet been measured at the LHC and RHIC.  Our study here shows that the energy, density and transverse momenta dependence of photon-jet (and photon-hadron) correlations both at near side and away side provide essential complementary information to understand the true underlying dynamics of ridge phenomenon in high-energy collisions. It was shown that the di-hadron ridge effect can be understood entirely within the framework of the initial-state CGC physics  \cite{ridge1,ridge2,ridge3,ridge4,ridge5}, see also Refs. \cite{ridge-out1,ridge-out2,v2-amir,energy-ridge}. There are also competing mechanisms based on the final state effects, like Hydrodynamics approach \cite{hydro-1,hydro-2}, which provide excellent description of the same data. Our study here may indicate that the ridge effect in high-multiplicity p+p(A) collisions is universal for different two-particle production, see also \cite{di-photon,di-jpsi}.  

Note that here we only considered prompt photon and quark pair production due to scatterings of a single quark (antiquarks) to the CGC target \cite{pho-cgc1,pho-cgc2,di-photon}. The actual cross section of photon-jet production also contains contribution of jets (quarks and gluons) and prompt photons produced independently from different quarks and (antiquarks). However, one may not expect any long-range correlated emissions from independent production. It will likely amount to an uncorrelated pedestal proportional to the product of single photon (or single jet) inclusive cross sections. Nevertheless, this effect remains to be investigated in details in future. 

The main building part of photon-jet production in electron-nucleus collisions is similar to the ones in quark-nucleus collisions at small-x. Therefore, our general outcome here can also be tested in a future electron-ion collider \cite{eic1,eic2}. A detailed study of photon-jet correlations in high-energy electron-nucleus collisions is postponed in future publication.

\begin{acknowledgments}
The author would like to thank Alex Kovner for useful discussions. 
This work is supported in part by Fondecyt grants 1110781.
\end{acknowledgments}



\begin{thebibliography}{99}

\bibitem{exp-pp}
V. Khachatryan {\it et al.} (CMS Collaboration), JHEP {\bf 09}, 091 (2010) [arXiv:1009.4122].
\bibitem{exp-pa1}
 S. Chatrchyan {\it et al.} (CMS Collaboration), Phys. Lett. {\bf B718}, 795 (2013) [arXiv:1210.5482].

\bibitem{exp-pa2}
 B. Abelev  {\it et al.}  (ALICE Collaboration), Phys. Lett. {\bf B719}, 29 (2013) [arXiv:1212.2001].

\bibitem{exp-pa3}
G. Aad  {\it et al.}  (ATLAS Collaboration), Phys. Rev. Lett. {\bf 110}, 182302 (2013) [arXiv:1212.5198].

\bibitem{exp-pa4}
B. B. Abelev  {\it et al.} (ALICE Collaboration), Phys. Lett. {\bf B726}, 164 (2013) [arXiv:1307.3237].

\bibitem{exp-pa5}
 A. Adare {\it et al.} (PHENIX Collaboration), Phys. Rev. Lett. {\bf 111}, 212301 (2013) [arXiv:1303.1794]; Phys. Rev. Lett. {\bf 114}, 192301 (2015) [arXiv:1404.7461].

\bibitem{exp-pa6}
L. Adamczyk {\it et al.} (STAR Collaboration), Phys. Lett. {\bf B743}, 333 (2015) [arXiv:1412.8437]; Phys. Lett. {\bf B747}, 265 (2015) [arXiv:1502.07652]. 


\bibitem{novel}
For example: V. Khachatryan {\it et al.} [CMS Collaboration], Phys. Rev. Lett. {\bf 106}, 122003 (2011); JHEP {\bf 1009}, 091 (2010); J. Adams {\it et al.} [STAR Collaboration], Phys. Rev. Lett. {\bf 91}, 072304 (2003); G. Aad {\it et al.} [ATLAS Collaboration], Phys. Rev. Lett. {\bf 105}, 252303 (2010); S. Chatrchyan {\it et al.} [CMS Collaboration], Phys. Rev. {\bf C84}, 024906 (2011); K. Aamodt {\it et al.} [ALICE Collaboration], Phys. Rev. Lett. {\bf 108} 092301 (2012). 


\bibitem{di-e}
A. Adare {\it et al.} [PHENIX Collaboration], Phys. Rev. Lett. {\bf 107}, 172301 (2011);
E. Braidot, for the STAR Collaboration, Nucl. Phys. {\bf A854}, 168 (2011); E. Braidot, Ph.D. thesis,  arXiv:1102.0931. 

\bibitem{ph-ex}
 A. Adare  {\it et al.} [PHENIX Collaboration], Phys. Rev. {\bf C80}, 024908 (2009). 

\bibitem{he-au}
 A. Adare {\it et al.} (PHENIX Collaboration), Phys. Rev. Lett. {\bf 115} (2015) 142301 [arXiv:1507.06273]. 


\bibitem{exp-hic}
S. Chatrchyan {\it et al.} (CMS Collaboration),  Phys. Lett. {\bf B724}, 213 (2013). 



\bibitem{exp-geo1}
B. Alver and G. Roland, Phys. Rev. {\bf C81}, 054905 (2010), erratum-ibid. {\bf C82}, 039903 (2010), 1003.0194. 
\bibitem{exp-geo2}
 L. Adamczyk {\it et al.} (STAR Collaboration), Phys. Rev. {\bf C88}, 014904 (2013) [arXiv:1301.2187].

\bibitem{hydro-1}
 P. Bozek, Eur. Phys. J. {\bf C71}, 1530 (2011) [1010.0405]; Phys. Rev. {\bf C88}, 014903 (2013);  P. Bozek and W. Broniowski, Phys. Lett. {\bf B718}, 1557 (2013) [arXiv:1211.0845].

\bibitem{hydro-2}
P. Bozek, W. Broniowski and G. Torrieri, Phys. Rev. Lett. {\bf 111}, 172303 (2013). 



\bibitem{ridge1}
A. Dumitru, K. Dusling, F. Gelis, J. Jalilian-Marian, T. Lappi and Venugopalan, Phys. Lett. {\bf B697}, 21 (2011) [arXiv:1009.5295]. 

\bibitem{ridge2}
A. Kovner and M. Lublinsky, Phys. Rev. {\bf D83}, 034017 (2011) [arXiv:1012.3398]; Phys. Rev. {\bf D84}, 094011 (2011) [arXiv:1109.0347].

\bibitem{ridge3}
E. Levin and A. H. Rezaeian, Phys. Rev. {\bf D84}, 034031 (2011) [arXiv:1105.3275].

\bibitem{ridge4}
K. Dusling and R. Venugopalan,  Phys. Rev. {\bf D87}, 094034 (2013);  {\bf D87}, 054014 (2013); {\bf D87}, 051502 (2013). 

\bibitem{ridge5}
T. Altinoluk, N. Armesto, G. Beuf, A. Kovner and M. Lublinsky, Phys. Lett. {\bf B751}, 448 (2015) [arXiv:1503.07126].  









\bibitem{ridge-rev1}
A. Kovner and M. Lublinsky, Int. J. Mod. Phys. {\bf E22}, 1330001 (2013) [arXiv:1211.1928]. 

\bibitem{ridge-rev2}
W. Li, Mod. Phys. Lett. {\bf A27}, 1230018 (2012) [arXiv:1206.0148].

\bibitem{ridge-rev3}
K. Dusling, W. Li and B. Schenke, Int.\ J.\ Mod.\ Phys.\  {\bf E25}, 1630002 (2016) [arXiv:1509.07939]. 

\bibitem{di-photon}
A. Kovner and A. H. Rezaeian, Phys. Rev. {\bf D90}, 014031 (2014) [arXiv:1404.5632]; Phys. Rev. {\bf D92}, 074045 (2015) [arXiv:1508.02412]. 




\bibitem{di-1} 
C.~Marquet,
  Nucl.\ Phys.\ {\bf A796}, 41 (2007)
  [arXiv:0708.0231]. 
\bibitem{di-2} 
  K. Tuchin, Nucl. Phys. {\bf A846}, 83 (2010). 
\bibitem{di-3} 
  J.~L.~Albacete and C.~Marquet,
  Phys.\ Rev.\ Lett.\  {\bf 105},162301 (2010).
\bibitem{di-4} 
F. Dominguez, B. W. Xiao and F. Yuan, Phys. Rev. Lett. 106, 022301 (2011);  F. Dominguez, C. Marquet, B.W. Xiao and F. Yuan, Phys. Rev.
{\bf D83}, 105005 (2011); B. W. Xiao and F. Yuan, Phys. Rev. {\bf D82}, 114009 (2010); Phys. Rev.
Lett. {\bf 105}, 062001 (2010); F. Dominguez, B.W. Xiao and F. Yuan, Phys. Rev. Lett. {\bf 106}, 022301 (2011) [arXiv:1009.2141]. 
\bibitem{di-5} 
A. Stasto, Bo-Wen Xiao and F. Yuan, Phys. Lett. {\bf B716},430 (2012) [arXiv:1109.1817]. 

\bibitem{dijet-diff1}
T. Altinoluk, N. Armesto, G. Beuf and A. H. Rezaeian, arXiv:1511.07452.
  \bibitem{dijet-diff2}
Y. Hatta, B-W Xiao and F. Yuan, arXiv:1601.01585. 


\bibitem{pho-cgc1} 
  F.~Gelis and J.~Jalilian-Marian,
  Phys.\ Rev.\  {\bf D66}, 014021 (2002). 
\bibitem{pho-cgc2} 
  R.~Baier, A.~H.~Mueller and D.~Schiff,
  Nucl.\ Phys.\  {\bf A741}, 358 (2004).

\bibitem{ja}
 J. Jalilian-Marian and A. H. Rezaeian, Phys. Rev. {\bf D86}, 034016 (2012) [arXiv:1204.1319].
\bibitem{ja2}
 A. H. Rezaeian, Phys. Rev. {\bf D86}, 094016 (2012) [arXiv:1209.0478].

\bibitem{jav1}
J. L. Albacete, N. Armesto, J. G. Milhano, P. Quiroga Arias and C. A. Salgado, Eur. Phys.  J. {\bf C71}, 1705 (2011). 
\bibitem{ip-sat}
A. H. Rezaeian, M. Siddikov, M. Van de Klundert and R. Venugopalan, Phys. Rev. {\bf D87}, 034002 (2013) [arXiv:1212.2974].
\bibitem{bcgc}  
A. H. Rezaeian and I. Schmidt, Phys. Rev. {\bf D88}, 074016 (2013) [arXiv:1307.0825].


\bibitem{ph-th}
X.-N. Wang and Z. Huang, Phys. Rev. {\bf C55}, 3047 (1997); X.-N. Wang, Z. Huang, and I. Sarcevic, Phys. Rev. Lett.
{\bf 77}, 231 (1996); H. Zhang, J. F. Owens, E. Wang and X.-N. Wang, Phys. Rev. Lett. {\bf 103}, 032302 (2009);
G.-Y. Qin, J. Ruppert, C. Gale, S. Jeon and G. D. Moore, Phys. Rev. {\bf C80}, 054909 (2009). 



\bibitem{sg}
L. V. Gribov, E. M. Levin and M. G. Ryskin, Phys. Rept. {\bf 100}, 1 (1983); A. H. Mueller and  J-W. Qiu, Nucl. Phys. {\bf 268}, 427
(1986).

\bibitem{mv} 
  L.~D.~McLerran and R.~Venugopalan,
  Phys.\ Rev.\ {\bf D49}, 2233 (1994); Phys. Rev. {\bf D49}, {\it ibid}. {\bf 49}, 3352 (1994); {\it ibid}. {\bf 50}, 2225 (1994).

\bibitem{bk}
I.~Balitsky,
Nucl.\ Phys.\  {\bf B463}, 99 (1996);
Y.~V.~Kovchegov,
Phys.\ Rev.\   {\bf D60}, 034008 (1999);
Phys.\ Rev.\   {\bf D61}, 074018 (2000).

\bibitem{jimwlk}
J. Jalilian-Marian, A. Kovner, A. Leonidov and H. Weigert, Nucl. Phys. {\bf B504}, 415 (1997); {\it ibid.}, Phys. Rev. {\bf D59}, 014014
(1999); E. Iancu, A. Leonidov and L. D. McLerran, Nucl. Phys. {\bf A692}, 583 (2001); E. Ferreiro, E. Iancu, A. Leonidov and L. D. McLerran, Nucl. Phys. {\bf A703}, 489 (2002). 

\bibitem{na-sat}
  N. Armesto and A.~H.~Rezaeian, Phys. Rev. {\bf D90}, 054003 (2014) [arXiv:1402.4831]. 

\bibitem{e-lhc}
CMS Collaboration, Phys. Rev. Lett. {\bf 105} 022002 (2010) [arXiv:1005.3299]; 
ALICE Collaboration, Phys. Rev. Lett. {\bf 105}, 252301 (2010) [arXiv:1011.3916];
CMS Collaboration, J. High Energy Phys. {\bf 08}, 141 (2011) [arXiv:1107.4800];  ATLAS Collaboration, Phys. Lett. {\bf B710}, 363 (2012) [arXiv:1108.6027].

\bibitem{m1}
E. Levin and A. H. Rezaeian, Phys. Rev. {\bf D82}, 014022 (2010) [arXiv:1005.0631]; Phys. Rev. {\bf D83}, 114001 (2011) [arXiv:1102.2385]; 
Phys. Rev. {\bf D82}, 054003 (2010) [arXiv:1007.2430];  arXiv:1011.3591. 

\bibitem{tr}
P. Tribedy and R. Venugopalan, Nucl. Phys. A850, 136-156 (2011) [Erratum-ibid. {\bf A859}, 185 (2011)];
A. Dumitru and Y. Nara, Phys. Rev. {\bf C85}, 034907 (2012).
\bibitem{tr1}
D. Kharzeev, E. Levin and M. Nardi, Nucl. Phys. {\bf A747}, 609 (2005). 
\bibitem{j1}
J. L. Albacete and A. Dumitru,  arXiv:1011.5161. 
\bibitem{me-pa}
A. H. Rezaeian, Phys. Rev. {\bf D85}, 014028 (2012) [arXiv:1111.2312];  Phys. Lett. {\bf B727}, 218 (2013) [arXiv:1308.4736]; arXiv:1208.0026; arXiv:1110.6642.  


\bibitem{cgc-review1}
E. Iancu, A. Leonidov and L. McLerran, hep-ph/0202270; E. Iancu and R. Venugopalan, hep-ph/0303204; F. Gelis, E. Iancu, J. Jalilian-Marian and R. Venugopalan, Ann. Rev. Part. Nucl Sci. {\bf 60}, 463 (2010).  
\bibitem{cgc-review2}
J. L. Albacete and C. Marquet, Prog. Part. Nucl. Phys. {\bf 76}, 1 (2014); J. L. Albacete, A. Dumitru and C. Marquet,  Int. J. Mod. Phys. {\bf A28}, 1340010 (2013). 

\bibitem{hybrid}  
A. Dumitru, A. Hayashigaki and J. Jalilian-Marian,
 Nucl. Phys. A {\bf 765}, 464 (2006); 
 G. A. Chirilli, Bo-Wen Xiao and Feng Yuan,  Phys. Rev. D {\bf 86}, 054005 (2012); T. Altinoluk, N. Armesto, G. Beuf, A. Kovner and  M. Lublinsky, arXiv:1411.2869. 

\bibitem{boris}
  B.~Z.~Kopeliovich, A.~V.~Tarasov and A.~Schafer,
  Phys.\ Rev.\  {\bf C59}, 1609 (1999)
  [hep-ph/9808378]. 

\bibitem{me2-pho}
A. H. Rezaeian and A. Schaefer, Phys. Rev. {\bf D81}, 114032  (2010) [arXiv:0908.3695]; B. Z. Kopeliovich, A. H. Rezaeian, H. J. Pirner and I. Schmidt, Phys. Lett. {\bf B653}, 210 (2007) [arXiv:0704.0642]; B. Z. Kopeliovich, E. Levin, A. H. Rezaeian and I. Schmidt, Phys. Lett. {\bf B675}, 190 (2009); B. Z. Kopeliovich, H. J. Pirner, A.H. Rezaeian, I. Schmidt, Phys. Rev. {\bf D77}, 034011 (2008); M. V. T. Machado and C. B. Mariotto, Eur. Phys. J. {\bf C61}, 871 (2009); E. Basso, V. P. Goncalves, M. Krelina, J. Nemchik and R. Pasechnik, arXiv:1603.01893.

\bibitem{nlo-bk1-2}
Y. V. Kovchegov and H. Weigert, Nucl. Phys. {\bf A784}, 188 (2007).







\bibitem{rcbk}
  J.~L.~Albacete and Y.~V.~Kovchegov,
  Phys.\ Rev.\  {\bf D75}, 125021 (2007).


\bibitem{bb}
I. I. Balitsky, Phys. Rev. {\bf D75}, 014001 (2007) [hep-ph/0609105].



\bibitem{me-jamal1}
  J. Jalilian-Marian and A. H. Rezaeian, Phys. Rev. {\bf D85}, 014017 (2012) [arXiv:1110.2810]. 

\bibitem{me-cgc}
A. H. Rezaeian, Phys. Lett. {\bf B718},1058 (2013) [arXiv:1210.2385]. 

\bibitem{raj}
K. Dusling, F. Gelis, T. Lappi and R. Venugopalan, Nucl. Phys. {\bf A836}, 159 (2010) [arXiv:0911.2720]. 

\bibitem{bk-b}
K. Golec-Biernat and A. M. Stasto, Nucl. Phys. {\bf B668}, 345 (2003); J. Berger and  A. M. Stasto, Phys. Rev. {\bf D84}, 094022 (2011); JHEP {\bf 1301}, 001 (2013) [arXiv:1205.2037].

\bibitem{urs}
N. Armesto, C. A. Salgado, and U. A. Wiedemann, Phys. Rev. Lett. {\bf 94},  022002 (2005). 


\bibitem{mstw}
A. D. Martin, W. J. Stirling, R. S. Thorne and G. Watt, Phys. Lett. {\bf B652}, 292 (2007);
A. D. Martin, W. J. Stirling, R. S. Thorne and G. Watt, Eur. Phys. J. {\bf C63}, 189 (2009). 

\bibitem{kkp}
B. A. Kniehl, G. Kramer and B. Potter, Nucl. Phys. {\bf B582}, 514 (2000).




\bibitem{fcc}
N. Armesto, A. Dainese, D. d'Enterria, S. Masciocchi, C. Roland, C.A. Salgado, M. van Leeuwen and U.A. Wiedemann, arXiv:1601.02963. 

\bibitem{dy-ana}
A. Stasto, B-W. Xiao and D. Zaslavsky, Phys. Rev. {D86}, 014009 (2012) [arXiv:1204.4861]. 


\bibitem{ridge-out1}
A. Dumitru and J. Jalilian-Marian, Phys. Rev. {\bf D81}, 094015 (2010) [arXiv:1001.4820]; 
E. Iancu and D. Triantafyllopoulos, JHEP {\bf 1111}  105 (2011) [arXiv:1109.0302]; 
  Y.~V.~Kovchegov and D.~E.~Wertepny, Nucl. Phys. {\bf A925},  254 (2014)
 [arXiv:1310.6701]. 

\bibitem{ridge-out2}
R. L. Ray, Phys. Rev. {\bf D84}, 034020  (2011) [arXiv:1106.5023]; Phys. Rev. {\bf D90}, 054013  (2014) [arXiv:1406.2736]; arXiv:1311.0774;
 B. Z. Kopeliovich, A. H. Rezaeian and I. Schmidt, Phys. Rev. {\bf D78}, 114009 (2008); B. Z. Kopeliovich, H. J. Pirner, A. H. Rezaeian and I. Schmidt, Phys. Rev. {\bf D77}, 034011 (2008). 











\bibitem{v2-amir}
A. Dumitru, L. McLerran and V. Skokov, Phys. Lett. {\bf B743}, 134 (2015). 
\bibitem{energy-ridge}
K Dusling, P. Tribedy and R. Venugopalan, Phys. Rev. {\bf D93}, 014034 (2016) [arXiv:1509.04410]. 



 \bibitem{di-jpsi}
S. P. Baranov and A. H. Rezaeian, arXiv:1511.04089. 

\bibitem{eic1}
E. Cartlidge, Nature {\bf 521}, 272 (2015). 
\bibitem{eic2}
D.~Boer, M.~Diehl, R.~Milner, R.~Venugopalan, W.~Vogelsang, D.~Kaplan, H.~Montgomery and S.~Vigdor {\it et al.},
  arXiv:1108.1713; A. Accardi  {\it et al.},  arXiv:1212.1701. 












\end{thebibliography}
\end{document}